\documentclass[11pt]{article}
\usepackage[margin=1in]{geometry}
\usepackage{amsmath,amssymb,booktabs,setspace,natbib}
\usepackage[T1]{fontenc}
\usepackage{lmodern}
\usepackage{graphicx}
\usepackage{longtable}
\usepackage{xcolor}

\usepackage{pdflscape}
\usepackage[colorlinks=true,urlcolor=blue,citecolor=blue,linkcolor=blue]{hyperref}
\title{Can Artificial Intelligence Forecast Treatment Effects in Economics?}
\author{Eva Vivalt\thanks{University of Toronto \href{mailto:eva.vivalt@utoronto.ca}{\texttt{eva.vivalt@utoronto.ca}}}
}
\date{November 20, 2026}

\begin{document}
\maketitle
\begin{abstract}
\noindent I use LLMs to forecast treatment effects from pre-registrations and pre-analysis plans, evaluating forecasts against results released after the models' cut-off dates. The screened sample includes 1,459 questions from 36 papers. For the most recent time period and model, the median absolute error is 0.053 SD, the average absolute error is 0.098 SD, and the MSE is 0.021, falling to 0.015 after adjusting for sampling variance where standard errors are available. Forecasts substantially outperform benchmarks predicting zero or other papers' mean effects. These results suggest considerable scope for forecasts to inform policy and research.

\end{abstract}


\clearpage
\doublespacing
\section{Introduction}

How well can we predict the results of a study before its findings are known? Decision-makers regularly draw on existing evidence to anticipate the effects of a treatment, but often high-quality evidence does not exist and they need to make their best guess. Large language models offer a way to generate predictions from descriptions of research projects before they resolve, however, the accuracy of these models is not known. Research on human forecasts documents they contain predictive power as well as systematic errors \citep{dellavigna2025,dreber2015,camerer2016}, and recent evidence establishes that AI can also predict effects in social science experiments \citep{ashokkumar2026}. This paper shows that LLMs can forecast the magnitudes of treatment effects across a broad range of economic interventions using pre-registrations and pre-analysis plans. Their forecasts are informative not only about expected effects, but also about uncertainty: predictive intervals have approximately their stated coverage, and wider intervals identify effects that are harder to predict.

This paper considers recent projects registered on the AEA RCT Registry and the Open Science Framework (OSF). The pre-registrations provide descriptions of planned trials in economics. I link these registrations to their earliest identified public results. I restrict attention to those with papers first shared in 2026, in order to use results released after the reported knowledge cutoff of GPT-5.5 in December 2025, and look at subsets from February 2026 and May 2026 in order to include still more recent models like Opus 4.8, Fable 5, and GPT-6 Astra. More capable models, as measured by the Epoch Capability Index (ECI), seem to do a better job forecasting than older models, though the sample sizes available after their knowledge cut-off dates are correspondingly smaller. In total, there were 3,068 forecasting targets for standardized treatment effects based on results from 39 papers. 1,459 of these, from 36 papers, passed some screening quality checks, and these form the main analysis sample.%

The median absolute forecast error for the most recent model (GPT-6 Astra, on a sample of 862 forecasts from 21 papers released since May 2026) is 0.053 SD, with a mean of 0.098 and a MSE of 0.021. For comparison, using the leave-one-paper-out mean within a time period to forecast effects in the remaining paper would produce a median absolute error of 0.082 SD, a mean of 0.122, and a MSE of 0.035. Another possible point of comparison is \cite{vivalt2020}, which collected results from 635 impact evaluations across 20 development interventions. Across 47 intervention-outcome combinations in that paper with at least 3 papers, a leave-one-out meta-analysis would predict the results of another study with a median absolute forecast error of 0.093 SD, mean of 0.170, and MSE of 0.095. While this sample of papers is different, and in principle a mixed model that included useful explanatory variables could do better at forecasting, this is nonetheless qualitatively a large gap in forecasting performance. AI has been doing better than humans on other types of forecasting tasks, such as forecasting geopolitical events \citep{fri2026}, and the models are only improving, so there seems to be considerable scope for AI forecasts to usefully inform decisions that would benefit from estimates of treatment effects.\footnote{RCT evidence may remain the gold standard, however, RCT evidence is frequently not available to decision-makers, so AI forecasts only need to improve on human forecasts in this case.}

A growing literature uses LLMs to simulate human responses and economic behavior \citep{argyle2023,horton2023}. \citet{ashokkumar2026} use simulated responses to predict experimental treatment effects, studying primarily survey experiments in psychology and political science, alongside a set of megastudies. \cite{Chetty2026} use ``digital twins'' to predict outcomes for workforce training participants in the absence of treatment. This paper examines forecasts of treatment effect magnitudes from pre-registration materials across a broad range of economic interventions.

This paper makes four main contributions. First, it shows that LLMs can directly forecast economic treatment effects with small errors across a range of interventions and settings. Second, it compares these forecasts against simple prediction benchmarks. Third, it illustrates that AI forecasts are not only good at estimating the mean effect, but are also pretty well-calibrated in their 10th–90th percentile predictive intervals and in ``knowing when they do not know'': wider intervals identify effects that are harder to predict. Finally, it shows that longer protocols are associated with more accurate forecasts.

These results imply that AI forecasts can be useful for informing priors in policy and research. These findings suggest that fuller study descriptions may help forecasting.
Finally, while this work suggests a role for forecasts, it also indirectly provides a perhaps sobering assessment of current economics research: if results can already be forecast so well, what is gained from research that only modestly overturns priors? Should we not be pursuing more ambitious projects?

\section{Data and forecast construction}
\label{sec:data}

I construct a dataset linking public descriptions of planned research with resultant reported treatment effects. Figure~\ref{fig:package-flow} summarizes the steps from identifying papers to constructing the evaluation sample.\footnote{Source documents, questions, extracted results, matching decisions and forecast responses underlying these steps are preserved in a replication package, though the responses to the prompts are non-deterministic.} The following sections describe each of these stages in more detail.

\subsection{Study identification and source materials}

I draw on three sources of study descriptions: public pre-analysis plans posted on the AEA RCT Registry, public pre-registrations without public pre-analysis plans posted on the AEA RCT Registry, and public pre-registrations on OSF. Registrations without a public pre-analysis plan are considered usable when their descriptions provide enough information for a LLM to formulate a numerical treatment effect forecasting question; beyond this criterion, I focus on those with particularly detailed descriptions. %

For the main analysis, I restrict attention to experiments whose earliest documented public findings are in 2026, based on systematic web searches, and record the date of first public release.\footnote{It is possible that some results may have made it into the training data for a particular model regardless. The extensive screening process is described in more detail later in the paper, but independent of paper screening, I do three things to mitigate this concern. First, papers where a large share of treatment effects are exactly or nearly-exactly estimated were closely inspected. For each result, a LLM considered if there was a plausible reason for the close match, such as the pre-analysis plan containing pilot results that were comparable to the later paper result. Second, I perform sensitivity analyses that drop each paper in turn. Third, Appendix Figure~\ref{fig:paper-removal} assesses sensitivity to progressively excluding the most accurately forecast papers, including excluding large shares of the corpus.} To search for results, registries are searched for linked results papers; targeted searches of working paper repositories are implemented, such as searches of NBER papers that first came out in 2026; and public materials on the internet are searched.

This approach identified 100 candidate records, of which 39 papers satisfied the paper-level requirements concerning when results were seemingly first made publicly available and whether pre-registration information was sufficiently detailed to enable forecasting treatment effects that could later be standardized.\footnote{Forecasts were elicited in raw units; information on standard deviations could come from either the pre-registration documents or the resultant paper.} The 39 candidates at this stage comprise 18 papers with public AEA pre-analysis plans, 18 with AEA registrations without public plans, and three with OSF pre-registrations.

The resultant sample is predominantly experimental. Of the 39 papers remaining at this stage, 37 use some form of randomized design. The two nonrandomized studies, both found in OSF, examine school phone bans and SNAP sugary drink restrictions using difference-in-differences designs. The predominance of randomized control trials (RCTs) in the corpus has several benefits. For example, it may mitigate the influence of selective reporting \citep{Brodeur2020, Vivalt2019}, and if a paper later reports results with slightly different controls than initially specified, it is unlikely to substantively affect treatment effect estimates.

These papers' protocols must still support sufficiently detailed questions that can be matched with results and standardized. The next section describes the construction of forecasting questions.

\subsection{Constructing forecasting questions}

The pre-registration and pre-analysis plan materials were used to construct forecasting questions without access to the resultant papers. Claude Opus 5 was used to initially propose questions, while GPT-6 Astra was used to review Claude's outputs for coverage, consistency, duplicate questions, units, and clarity. These models were instructed to construct and review questions using the supplied registration and pre-analysis plan materials. They were not supplied resultant papers and were not informed of the accuracy metric that would later be used to evaluate the forecasts. The research package retains the prompts, submitted materials and responses.

Questions specify the intervention, treatment arm, outcome, sample, time period, unit, and specification (if provided in the pre-registration documents). Questions were not restricted to primary outcomes but also include heterogeneous treatment effects and results reported at different time periods. If a pre-registration is unclear on some dimensions, such as not reporting the time period, it is not excluded on that basis, however, each question requires an identifiable intervention or treatment contrast, substantive outcome, relevant population, and numerical scale. Explicit conventions were used to determine how to treat questions that were almost fully specified but for some detail. Appendix Figure~\ref{fig:forecast-question-examples} provides an example of a resulting question.

Using these broad criteria, 14,103 questions were extracted from 39 papers. Most, however, could not be connected to unique numerical results in the papers, as described in the next section.

\subsection{Extracting and matching reported results}

A separate numerical extraction step extracted values from the identified papers. The intervention, treatment arm, outcome, sample, time period, unit and specification, where available, were logged and the closest match selected between the extracted results and each forecasting question. Forecasting questions and results were matched 1:1, with duplicates removed. Exact matches in the content between a forecasting question and a result were distinguished from close but inexact matches. If a result was reported in a way that flipped the sign from the original forecasting question, that was also identified and results flipped before analysis. Results were extracted in the way they were reported (e.g., a treatment coefficient and standard error vs. a treatment group mean, standard deviation, and number of observations, along with a control group mean, standard deviation, and number of observations). Results were also read off of figures, where applicable, and a flag was added to indicate if a result was extracted in this way.

Additional review rounds checked the identified results against the supplied papers and assessed correspondence to the forecasting question. A ``match score'', on a 0--100 scale, provides a continuous measure of the degree of match between the description in the pre-registration and the paper for each specific question. The match score review and the two eligibility assessments were both blinded to the models' returned forecasts and forecast errors; forecasting questions were compared against the paper results without having the forecasts to refer to.

Overall, 5,307 matched question--result pairs were produced from 39 papers.

\subsection{Reporting units and standardization}

Forecasts were elicited in ``raw'' units where feasible and converted to comparable standardized effects. A compatible baseline outcome standard deviation was preferred for standardizing where available; otherwise, a documented compatible control group, pooled, follow-up, or other outcome SD was used, including in a few instances historical SDs provided in the pre-registration documents or SDs reconstructed under assumptions.\footnote{Such as SDs reconstructed as standard errors $\times$ root-$N$ and SDs reconstructed using a regression $R^2$ identity. Results are similar when questions requiring reconstructed SDs are excluded.} For binary outcomes, the SD was constructed from reference probabilities.

Of the 5,307 matched pairs from 39 papers, 3,171 questions from 39 papers yielded usable standardized results.

\subsection{Forecast elicitation and eligibility}

Forecasting requests contain the study protocols, forecasting questions and reporting instructions, but no results, and are implemented through the API as a closed-book exercise without any web search/browsing/retrieval/or coding tools enabled. The prompts request a predictive mean and distributional quantiles and explicitly incorporate both uncertainty about the effect and sampling variability. The main analyses use xhigh reasoning effort for GPT-5.5 and GPT-6 Astra and max effort for the Anthropic models, with a 64,000-token output cap. GPT-5.1 uses high, its maximum available effort, with a 32,000-token cap (Appendix Table~\ref{tab:higher-effort}). Additional checks assessed that the returned forecasts were consistent with the requested units and time aggregation, using only the forecasting inputs and returned forecasts and explanations, without access to the realized results or forecast errors. When this information established that a forecast should be rescaled or converted, that unique conversion was applied.

This paper uses GPT-5.1, GPT-5.5, Opus 4.6, Opus 4.8, Fable 5 and GPT-6 Astra. Papers from January onwards are considered for GPT-5.1, GPT-5.5 and Opus 4.6, papers from February onwards are used for Opus 4.8 and Fable 5, and papers from May for Astra, per provider-reported model cutoffs (Table~\ref{tab:model-cutoffs}).

For the analysis, a question must have a match score of at least 60 and neither of the two eligibility assessments may identify a demonstrated material mismatch. A material mismatch requires evidence that the result answers a substantively different question regarding the intervention, treatment arm, outcome, sample, time period, or unit.\footnote{In addition, the correspondence between the question and result may be considered to have a minor discrepancy that does not meet this standard, or be unresolved. Those classifications do not disqualify a potential question.}

The set of 39 studies excludes papers with numerical or qualitative findings from the same experiment publicly documented before the start of the model's evaluation window. During forecasting, models also flagged if they recalled any results. No results were flagged for these 39 papers, but this additional screen could have been valuable: the models do flag some results for papers excluded for having had some past public disclosures. Of the 3,171 matched question-result pairs that could be standardized, 103 were excluded for representing non-treatment quantities, such as results from balance tables, leaving 3,068 forecasting targets; the correspondence screen then excludes 1,549. I additionally exclude 60 questions concerning interactions with continuous moderators from the main analysis; restoring them leaves the MSEs virtually unchanged.\footnote{Results available upon request.} Applying these criteria leaves 1,459 questions from 36 papers in January 2026 or later (``Jan+''), 1,386 from 32 in Feb+, and 862 from 21 in May+, as shown in Figure~\ref{fig:package-flow}. Table \ref{tab:source-counts} provides further details on the resultant set of papers and forecasting questions and their sources. The 36 papers in the January-onward screened sample include 16 with public AEA RCT Registry pre-analysis plans, 17 with AEA RCT Registry pre-registrations without public pre-analysis plans, and 3 with OSF pre-registrations. The papers retained in the screened analysis sample are listed in Appendix~\ref{app:included-papers}.

To assess accuracy, I calculate squared errors in SD units, average within each paper, and then weight papers equally. Neither results nor forecasts are winsorized. The 10th and 90th percentiles were also elicited, defining an 80\% predictive interval.

\section{Results}
\label{sec:main-results}

\subsection{The distribution of treatment effects and forecasts}
\begin{figure}[p]\centering
\includegraphics[width=\linewidth]{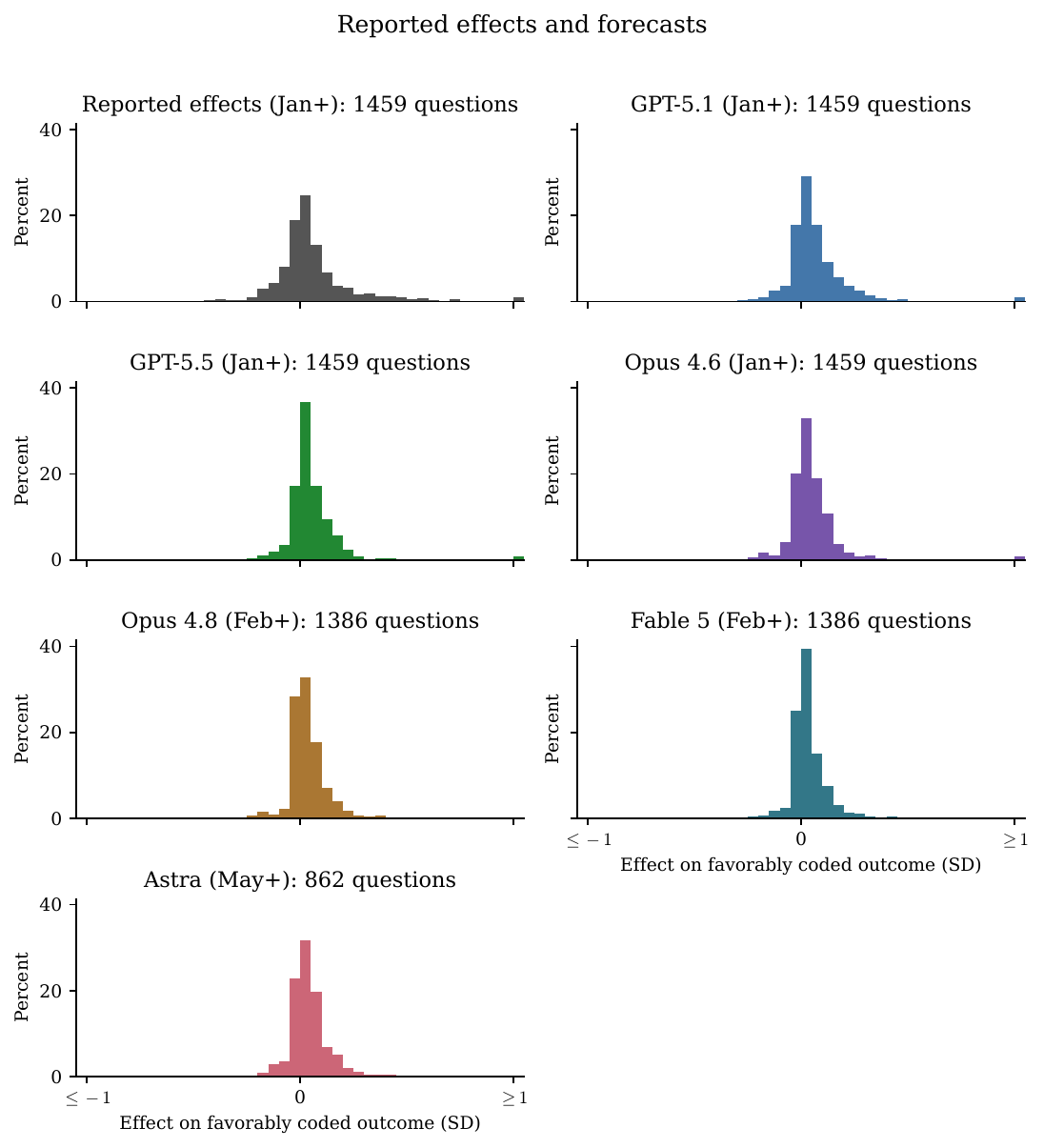}
\caption{Reported effects and forecasts}
\label{fig:signed-jan-combined}
\par
\begin{minipage}{0.97\linewidth}\footnotesize Outcomes are oriented so that an increase is considered a positive outcome. Each question has equal weight. The tails are binned in end bins.\end{minipage}
\end{figure}

Figure \ref{fig:signed-jan-combined} plots distributions of reported treatment effects and forecasts. Each question receives equal weight. For the sake of this figure, questions are oriented so that a positive treatment effect represents a good outcome. There is a long right-hand tail of realized treatment effects and relatively few negative results or forecasts of negative results.

\subsection{Forecast accuracy}
Table \ref{tab:accuracy} reports the MSE by model and time period. Results for older models are also shown in the Feb+ and May+ subset to enable closer comparison with models only eligible beginning in those time periods, since the samples otherwise differ. MSEs are generally small and declining for more recent models. Appendix Table~\ref{tab:accuracy-question-weights} reports MSEs weighting questions equally rather than papers equally. Panel B shows results on the subset of questions that could be adjusted for sampling variance using
\begin{equation}
\mathrm{MSE}_{\mathrm{adjusted}}=\sum_i w_i\left[(f_i-\widehat{\tau}_i)^2-\mathrm{SE}_i^2\right],
\label{eq:sampling-variance-adjustment}
\end{equation}
where $f_i$ is the forecast, $\widehat{\tau}_i$ is the reported effect, and the weights $w_i$ sum to one, giving papers equal weight and questions equal weight within papers. As this calculation requires standard errors to be available or recoverable, the sample is smaller; a version of Panel A on this subset of forecasting questions is provided in Appendix Table~\ref{tab:accuracy-se-available} for comparison.

Table \ref{tab:accuracy} also shows that the AI forecasts perform well compared to benchmarks of always predicting 0 as a treatment effect or predicting the leave-one-paper-out mean within the sample.\footnote{This benchmark uses realized results, rather than only information available when a forecast could have been made, which may advantage it if treatment effects change over time. Effects retain the directions specified by the forecasting questions. Using the favorable-outcome coding yields similar unadjusted benchmark MSEs: 0.075, 0.047 and 0.035 for Jan+, Feb+ and May+, respectively.} To further benchmark these results, I compare forecasts against the ``best guess'' that one might make through meta-analysis. In particular, I use the data set in \cite{vivalt2020}, formed from a harmonized set of 20 meta-analyses and systematic reviews, which cover 47 intervention-outcome combinations (e.g., the effects of unconditional cash transfers on enrollment rates). For each intervention-outcome combination, I repeatedly drop one paper and see what the MSE would be if it had been estimated by all the other papers in the set using an inverse-variance-weighted random-effects meta-analysis. Appendix Tables~\ref{tab:jeea-mse-tests} and \ref{tab:jeea-mse-tests-five} report the historical benchmark comparisons. There are some differences between the samples, so results are not directly comparable; Appendix Figure~\ref{fig:effect-histogram} compares the magnitudes of the reported effects in the current and historical samples.

\begin{table}[p]\centering\small
\caption{Forecast MSE by model and period}
\label{tab:accuracy}
\begin{tabular}{lrrr}
\toprule
 & Jan+ & Feb+ & May+ \\ \midrule
\multicolumn{4}{l}{\textit{Panel A. Unadjusted MSE (full sample)}} \\
\addlinespace[3pt]
\multicolumn{4}{l}{\textit{AI models}} \\
GPT-5.1 & 0.0398 & 0.0397 & 0.0299 \\
GPT-5.5 & 0.0323 & 0.0328 & 0.0233 \\
Opus 4.6 & 0.0327 & 0.0332 & 0.0226 \\
Opus 4.8 & --- & 0.0323 & 0.0226 \\
Fable 5 & --- & 0.0323 & 0.0235 \\
Astra & --- & --- & 0.0213 \\
\midrule
\multicolumn{4}{l}{\textit{Benchmarks}} \\
Always predict zero & 0.0856 & 0.0541 & 0.0400 \\
Leave-one-paper-out mean & 0.0745 & 0.0465 & 0.0348 \\
\midrule
Papers & 36 & 32 & 21 \\
Questions & 1459 & 1386 & 862 \\
\midrule
\multicolumn{4}{l}{\textit{Panel B. Sampling-variance-adjusted MSE (SE available)}} \\
\addlinespace[3pt]
\multicolumn{4}{l}{\textit{AI models}} \\
GPT-5.1 & 0.0288 & 0.0285 & 0.0245 \\
GPT-5.5 & 0.0199 & 0.0202 & 0.0162 \\
Opus 4.6 & 0.0204 & 0.0207 & 0.0149 \\
Opus 4.8 & --- & 0.0187 & 0.0148 \\
Fable 5 & --- & 0.0199 & 0.0160 \\
Astra & --- & --- & 0.0145 \\
\midrule
\multicolumn{4}{l}{\textit{Benchmarks}} \\
Always predict zero & 0.0856 & 0.0475 & 0.0383 \\
Leave-one-paper-out mean & 0.0713 & 0.0375 & 0.0323 \\
\midrule
Papers & 31 & 27 & 19 \\
Questions & 934 & 871 & 538 \\
\bottomrule\end{tabular}
\par\vspace{7pt}\begin{minipage}{0.97\linewidth}\footnotesize Errors are averaged within paper, then equally across papers in each column. Benchmarks are calculated within each time period's sample. Panel B shows results for questions with an available standard error after subtracting off sampling variance.\end{minipage}\end{table}

\clearpage
\begin{figure}[p]\centering
\includegraphics[width=\linewidth]{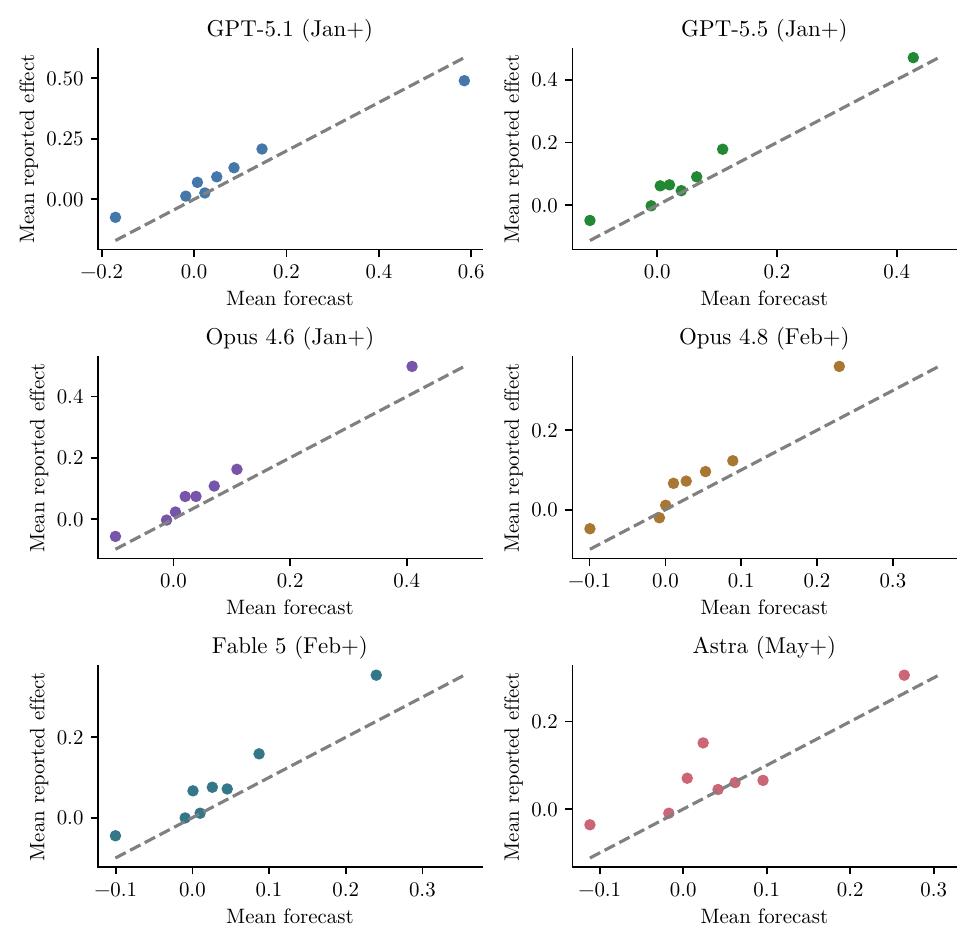}
\caption{Binned calibration by model}
\label{fig:diagnostic-calibration}
\par
\begin{minipage}{0.97\linewidth}\footnotesize This figure plots the mean forecast against the mean reported effect, both in standard deviations, for each model. Different models use different samples given their different cut-off dates.\end{minipage}
\end{figure}

Appendix Figure~\ref{fig:diagnostic-cdf} shows the distribution of absolute forecast errors. Appendix Figure~\ref{fig:mse-eci} plots the MSE of different models against their Epoch Capability Index. The pattern suggests that more capable models tend to forecast more accurately, although the comparison includes only a few models and the differences are not always statistically significant; the largest difference is between GPT-5.1 and the other models.

Figure~\ref{fig:diagnostic-calibration} compares mean reported effects with mean forecasts across bins of forecast magnitude. Forecasts appear to be relatively well-calibrated: mean reported effects generally increase with mean forecasts. Appendix Figure~\ref{fig:predictability-uncapped} provides more detail, showing individual forecasts and estimating a line of best fit. On Opus 4.8, Fable 5 and Astra, for example, the slopes are 0.86, 0.90 and 0.86 on the May+ sample. 

Appendix Figure~\ref{fig:diagnostic-papers} plots the MSE by paper. The values are clearly driven by a few outlier papers. This is potentially important if one were concerned about study leakage of results to the models; leakage would presumably mean some papers had overly small MSEs, but the aggregate MSE would not move much if excluding a paper that was accurately forecast because of leakage and would plausibly move much more if dropping a paper with a large MSE. Appendix Figure~\ref{fig:paper-inclusion-combined} assesses the sensitivity of these results to paper inclusion, and they seem relatively robust. For Astra in May+, omitting one paper at a time gives MSEs between 0.018 and 0.022, compared with 0.021 in the full sample. Appendix Figure~\ref{fig:paper-removal} further shows that each model's MSE remains below its leave-one-out mean benchmark when progressively removing its most accurately forecast papers.\footnote{The benchmark value is recalculated on the remaining papers after each removal. Different models can remove papers in different orders given their respective forecasts.} That is to say, even if the most accurately forecast papers were treated as potentially contaminated and excluded, the forecasts on the remainder would still outperform the benchmark. %

Re-running the forecasting prompt yields forecasts correlating 97\% across identical requests on average (Table~\ref{tab:forecast-repeatability}). I also assess sensitivity to different prompts at the same effort settings, as shown in Appendix Figure~\ref{fig:prompt-mse-changes-all-models}, but results do not change much. %

\clearpage

\subsection{Distributional results}
\begin{figure}[p]\centering
\includegraphics[width=\linewidth]{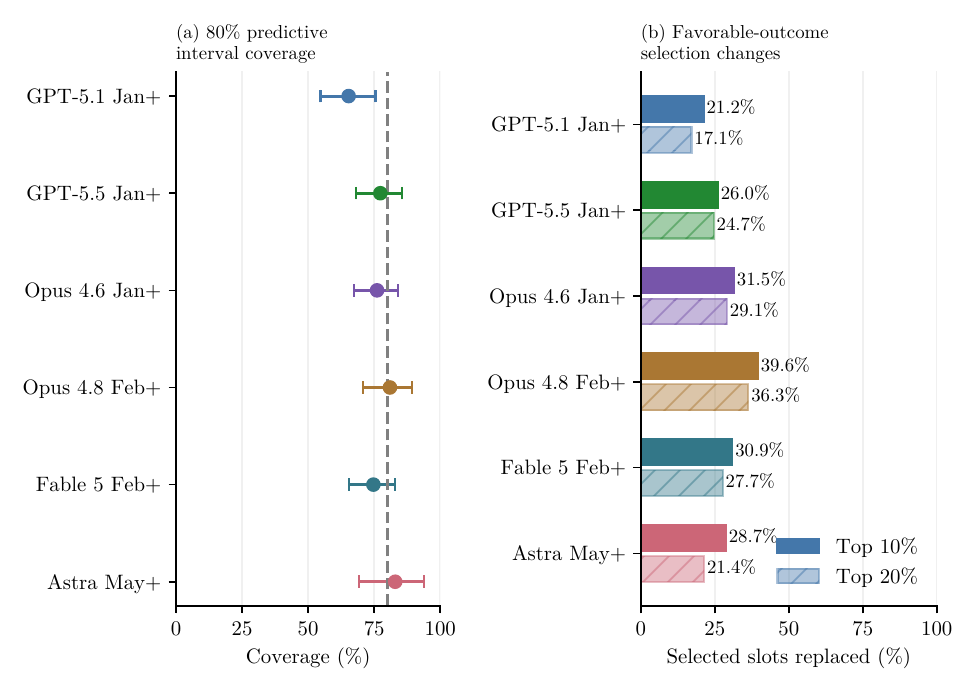}
\caption{Predictive interval coverage and changes in top-decile and top-quintile selection}
\label{fig:diagnostic-coverage}
\par
\begin{minipage}{0.97\linewidth}\footnotesize
Panel (a) shows inclusive coverage of the 10th--90th percentile interval against reported treatment effects, with equal paper weights and 95\% intervals from 2,000 bootstrap draws. The dashed line represents 80\% coverage. Panel (b) ranks questions by their mean or 90th-percentile forecast in SD units and selects the top 10\% or 20\% under each ranking. Bars show the percentage of selected questions that would be replaced when moving from a selection rule based on mean forecasts to a rule based on 90th-percentile forecasts. Jan+ includes 1,459 questions in 36 papers; Feb+ includes 1,386 in 32; May+ includes 862 in 21.
\end{minipage}\end{figure}

Figure \ref{fig:diagnostic-coverage} evaluates the 10th--90th percentile intervals against the reported treatment effects. If the models are well-calibrated about their uncertainty, results should fall within this interval 80\% of the time. In practice, the treatment effects fall within this interval 74.7--82.9\% of the time for five models, and 65.3\% for GPT-5.1; one can only reject that they fall within the interval 80\% of the time for GPT-5.1. The second panel of Figure~\ref{fig:diagnostic-coverage} illustrates how rankings differ when using forecast quantiles rather than means. Questions' rank according to their 90th percentile estimate has a Spearman correlation of 0.73--0.86 with their rank at the mean, and for a decision-maker trying to select the top 10\%, judging by the 90th percentile rather than the mean would change 21.2--39.6\% of selections.\footnote{This abstracts from any differences in the costs of different interventions or value of different outcomes, considering all outcomes in SD units equally.} These rankings concern reported estimates, so a higher upper quantile can reflect greater sampling variability as well as the possibility of a larger underlying effect, but results still suggest some scope for distributional forecasts to inform policy decisions.

\begin{figure}[p]\centering
\includegraphics[width=\linewidth]{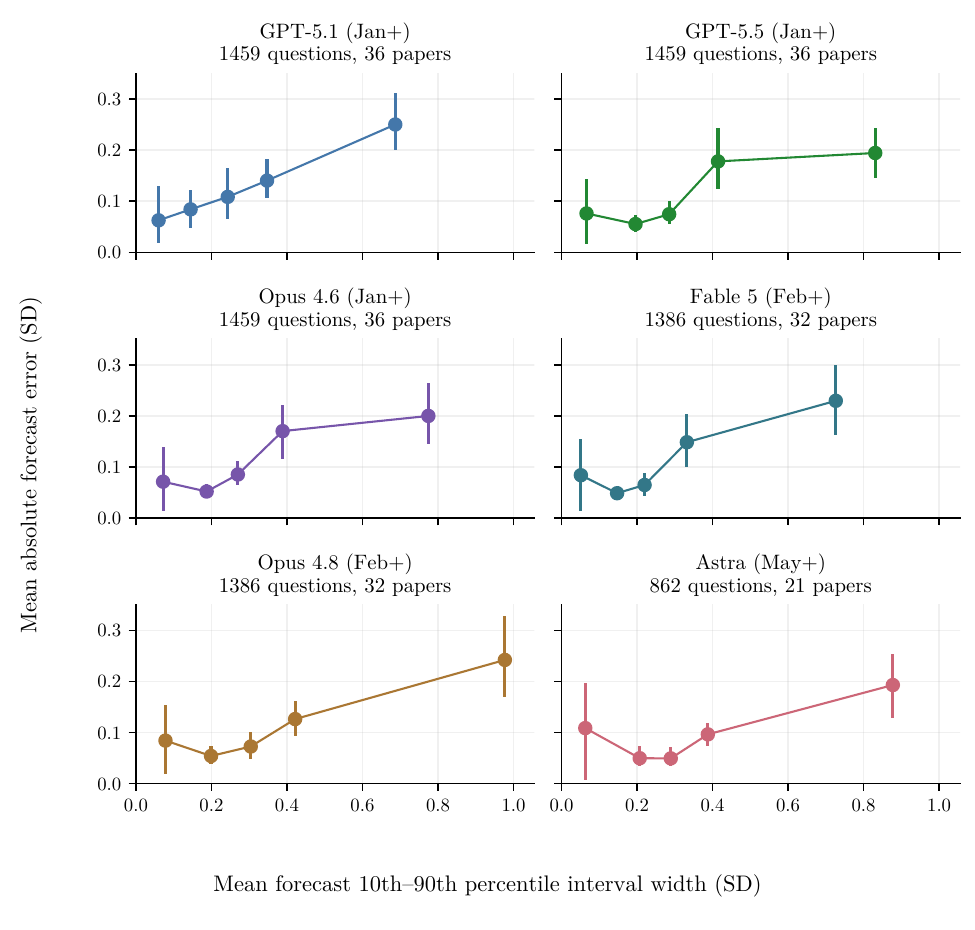}
\caption{Do wider forecast intervals identify less predictable results?}
\label{fig:uncertainty-error}
\par
\begin{minipage}{0.97\linewidth}\footnotesize
Questions are binned into five groups of forecast interval width. Questions are weighted equally within a paper and papers are equally weighted before binning. Vertical bars represent 95\% confidence intervals from 2,000 bootstrap draws. Errors are absolute differences between the mean forecast and reported effect, in standard deviations. Models' eligible windows differ; comparisons across panels are descriptive. Those with wider interval bins, indicating more uncertainty, appear to have larger forecast errors.\end{minipage}
\end{figure}

Figure~\ref{fig:uncertainty-error} asks whether wider forecast intervals identify less predictable results. Wider intervals do appear to generally accompany larger absolute forecast errors. This implies that models can be used to gauge which results are less predictable, beyond achieving coverage on average. This uncertainty concerns reported estimates and can reflect both uncertainty about the underlying treatment effects and sampling variability.

\clearpage
\section{Discussion and Limitations}
\label{sec:discussion}

The results describe how accurately LLMs can predict treatment effects from registrations and pre-analysis plans and seem to suggest substantial scope for these forecasts to improve decision-making.

Several limitations remain. One potential concern is that the models might have been aware of some results despite the use of training or knowledge cut-off dates. While this cannot be fully ruled out, several steps were taken to mitigate this issue. First, the LLMs were asked to provide closed-book forecasts through the API without access to the internet or any other tools while forecasting. Second, there were three stages at which papers and results would have been screened: (1) at the initial search for candidate records, which identified papers first released after the relevant cut-off dates but did not consider whether particular results within the paper may have been previously disclosed; (2) as an additional screening step, which searched for disclosure of specific results in companion papers, conferences or workshops, seminars, abstracts on an author's webpage or academic CV, or media citations; (3) as part of the forecasting exercise, in which models were asked to report when they recalled any results being forecast, as they might do if those results were in their training data. No results were flagged at stage (3) for the 39 papers in the corpus, but models did flag recalled results for papers excluded because of prior public disclosures.\footnote{Additionally, those with exact or particularly close matches were inspected post-hoc to sanity check if they seemed to have a reason for being so close, such as if a pre-analysis plan cited a pilot result which was similar to the result later reported in the paper.} Concerns may also be mitigated by the fact that different models from different companies would have to simultaneously not flag a disclosed result in order for it to pass screening. Finally, I assess the sensitivity of the results to dropping studies predicted with low MSEs: one can drop a fairly large share of studies and still have results that are qualitatively similar (Appendix Figure~\ref{fig:paper-removal}).\footnote{Appendix~\ref{app:disclosure-channels} provides further detail on the channels through which papers came to be excluded for potential disclosure.}

The screening rule considers a paper contaminated if there is a disclosure flag for any outcome in the paper, conservatively. A source audit found no main study effect findings in the materials supplied to forecasters, and question construction was restricted to the supplied registration and protocol materials. These safeguards constrain potential contamination via inputs or forecasting questions, though they cannot fully rule out results-informed protocol revisions. Appendix Table~\ref{tab:source-exclusions} provides robustness checks using alternative screening rules that additionally remove papers that report pilot findings, main study statistics such as take-up rates, or hypothetical results. Results look similar without these papers.

Importantly, the sample is selected in ways that could affect results. The findings apply only to outcomes with reported results that can be standardized and matched sufficiently closely to the pre-registered questions, and the matching of results to forecasting questions could introduce noise. Further, the focus is on studies with well-documented pre-registrations or pre-analysis plans, but these may be precisely the easiest to forecast. Indeed, studies with longer descriptions (as measured by number of words) tend to be forecast better after accounting for registration source, although the relationship varies across models and source categories (Figure~\ref{fig:protocol-length-coefficients}).\footnote{There are too few OSF papers to assess this relationship reliably.} Figure~\ref{fig:protocol-length} illustrates.

Rather than this latter issue being a drawback, however, it highlights that forecasts could do even better in the future if researchers wrote fuller descriptions of their studies, although differences between studies could also explain the relationship. With LLMs being used more in research, writing more complete descriptions may become easier. LLMs might both write scientific protocols and assess their treatments' likely effects.

Despite these limitations, it seems clear that forecasting can already serve as a useful tool - not just for forecasting states of the world like who wins an election or sports game, but also for forecasting causal treatment effects.

Forecasting treatment effects may at first glance seem like a step backwards: the credibility revolution occurred in large part to mitigate the risks of biases that can result in the absence of rigorous evidence. However, given we do not always have results from a RCT, and sometimes one cannot be done, high-quality forecasts may improve over the counterfactual of less well-informed and biased individual beliefs. Further, by repeatedly forecasting a variety of research results in different contexts, we can hope to learn more about the types of situations under which forecasts perform well. The evidence so far seems to suggest they are generally very accurate, and wider forecast intervals identify effects that are harder to predict. This suggests it may be worth focusing research efforts more on topics where there is more uncertainty in the AI forecasts, particularly where resolving that uncertainty could change a policy decision. Leveraging AI forecasts to inform experimental design could help researchers obtain more value from their studies.

\clearpage
\setstretch{1.12}

\clearpage\appendix
\counterwithin{figure}{section}
\counterwithin{table}{section}
\renewcommand{\theHfigure}{appendix.\thesection.\arabic{figure}}
\renewcommand{\theHtable}{appendix.\thesection.\arabic{table}}
\section{Supplementary tables and figures}
\begin{figure}[p]\centering

\includegraphics[width=\linewidth]{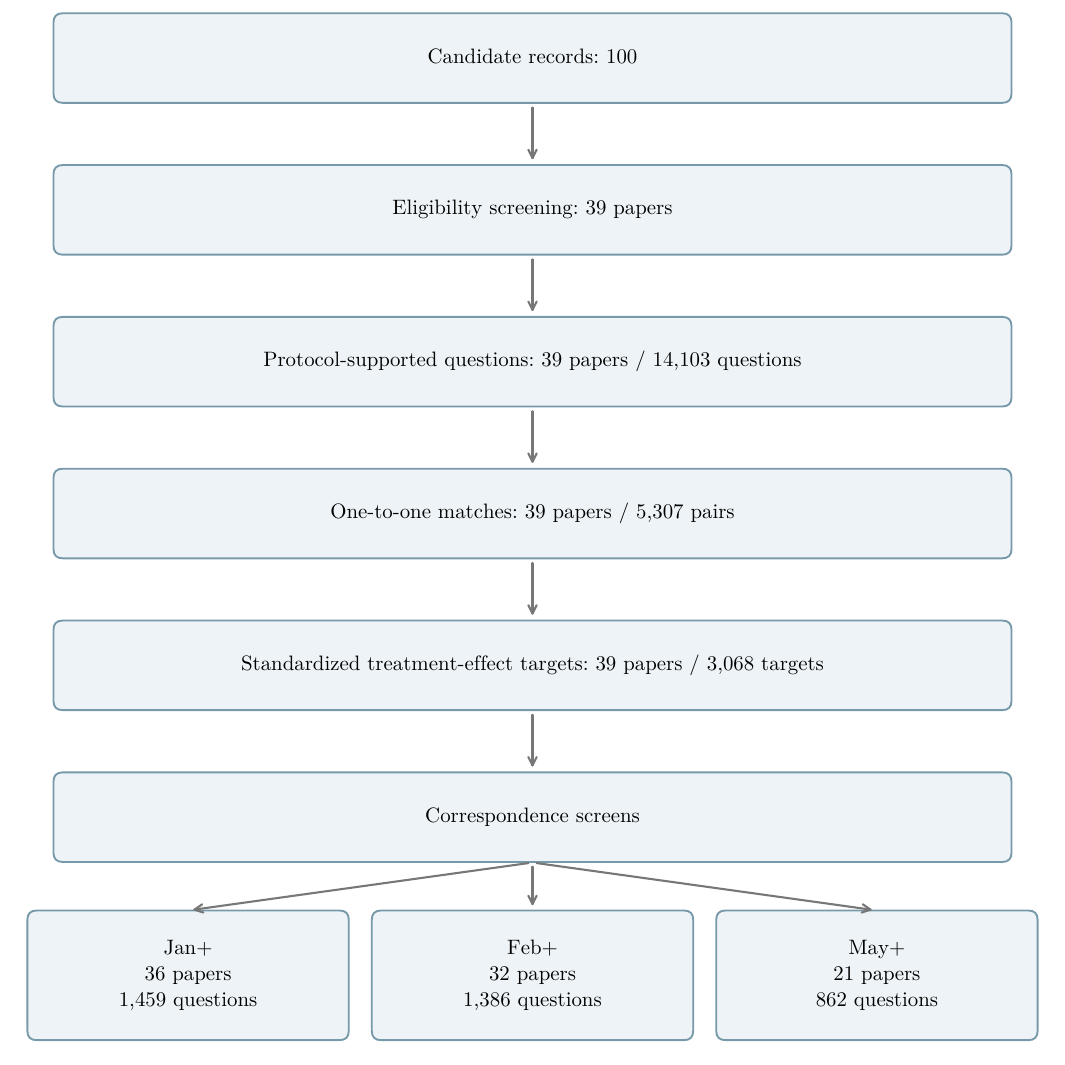}
\caption{Sample construction}
\label{fig:package-flow}
\par
\begin{minipage}{0.97\linewidth}\footnotesize This flowchart shows how papers and questions were screened at each stage.\end{minipage}
\end{figure}

\clearpage
\begin{figure}[p]
\centering
\setlength{\fboxsep}{10pt}
\fbox{\begin{minipage}{\dimexpr\linewidth-2\fboxsep-2\fboxrule\relax}
\small
\textbf{Nonlinear Pricing and Revealed Salience}
\medskip
\par\textbf{Question}\par
What will be the ITT effect of the pooled MOV arms (E.1, E.2, F.1, F.2, G.1, G.2, H.1, H.2) relative to control Group I on the probability that a randomized visitor purchases, in percentage points?
\par\medskip
\textit{Reporting instructions:} Return the numerical effect in percentage points. Preserve the outcome, treatment ordering, subgroup, measurement period and estimator in the target description. This reporting-unit instruction governs the answer unit; raw-scale descriptions elsewhere define measurement only. Do not change the substantive target to make a conversion easier.
\par\medskip\textbf{Outcome}\par
Probability to purchase at the shop
\par\medskip\textbf{Unit}\par
percentage points
\end{minipage}}
\caption{Example of a question submitted to the forecasting models. In addition, the forecasting models received study and question identifiers, protocol reference material, and common forecasting instructions.}
\label{fig:forecast-question-examples}
\end{figure}

\clearpage
\begin{table}[!ht]
\centering
\small
\caption{Provider-reported model cutoffs and evaluation periods}
\label{tab:model-cutoffs}
\begin{tabular}{@{}lp{3.2cm}p{4.5cm}l@{}}
\toprule
Model & Reported cutoff & Provider terminology & Evaluation period \\
\midrule
GPT-5.1
& September 30, 2024
& Knowledge cutoff
& Jan+ 2026 \\[4pt]

GPT-5.5
& December 1, 2025
& Knowledge cutoff
& Jan+ 2026 \\[4pt]

Opus 4.6
& May 2025; August 2025
& Reliable knowledge cutoff; training data cutoff, respectively
& Jan+ 2026 \\[4pt]

Opus 4.8
& January 2026
& Reliable knowledge cutoff and training data cutoff
& Feb+ 2026 \\[4pt]

Fable 5
& January 2026
& Reliable knowledge cutoff and training data cutoff
& Feb+ 2026 \\[4pt]

GPT-6 Astra
& April 30, 2026
& Knowledge cutoff
& May+ 2026 \\
\bottomrule
\end{tabular}

\par\vspace{7pt}
\begin{minipage}{0.97\linewidth}
\footnotesize
Reported cutoff and provider terminology come from Provider documentation for
\href{https://developers.openai.com/api/docs/models/gpt-5.1}{GPT-5.1},
\href{https://developers.openai.com/api/docs/models/gpt-5.5}{GPT-5.5},
\href{https://platform.claude.com/docs/en/models/opus-4-6/overview}{Opus 4.6},
\href{https://platform.claude.com/docs/en/models/opus-4-8/overview}{Opus 4.8},
\href{https://platform.claude.com/docs/en/models/fable-5/overview}{Fable 5},
and
\href{https://developers.openai.com/api/docs/models/gpt-6-astra}{GPT-6 Astra},
accessed October 18, 2026. The ``evaluation period'' refers to the window over which each model was considered eligible in this paper.
\end{minipage}
\end{table}

\clearpage
\begin{table}[p]\centering\small
\caption{Analysis sample by source}
\label{tab:source-counts}
\setlength{\tabcolsep}{4pt}
\begin{tabular}{lrrrrrr}
\toprule
Source & \multicolumn{2}{c}{Jan+} & \multicolumn{2}{c}{Feb+} & \multicolumn{2}{c}{May+} \\
\cmidrule(lr){2-3}\cmidrule(lr){4-5}\cmidrule(lr){6-7}
 & Papers & Questions & Papers & Questions & Papers & Questions \\ \midrule
AEA public PAP & 16 & 918 & 14 & 853 & 8 & 422 \\
AEA registration, no public PAP & 17 & 488 & 15 & 480 & 11 & 422 \\
OSF & 3 & 53 & 3 & 53 & 2 & 18 \\
\midrule
Total & 36 & 1459 & 32 & 1386 & 21 & 862 \\
\bottomrule\end{tabular}
\par\vspace{7pt}\begin{minipage}{0.97\linewidth}\footnotesize This table summarizes the number of papers and questions available at each time period from each source.
\end{minipage}\end{table}

\clearpage
\begin{table}[p]\centering\small
\caption{Forecast MSE with equal weights for questions}
\label{tab:accuracy-question-weights}
\begin{tabular}{lrrr}
\toprule
 & Jan+ & Feb+ & May+ \\
\midrule
\multicolumn{4}{l}{\textit{Panel A. Unadjusted MSE (full sample)}} \\
\addlinespace[3pt]
\multicolumn{4}{l}{\textit{AI models}} \\
GPT-5.1 & 0.0340 & 0.0318 & 0.0244 \\
GPT-5.5 & 0.0258 & 0.0240 & 0.0176 \\
Opus 4.6 & 0.0277 & 0.0258 & 0.0180 \\
Opus 4.8 & --- & 0.0244 & 0.0177 \\
Fable 5 & --- & 0.0243 & 0.0166 \\
Astra & --- & --- & 0.0162 \\
\midrule
\multicolumn{4}{l}{\textit{Benchmarks}} \\
Always predict zero & 0.0511 & 0.0336 & 0.0255 \\
Leave-one-paper-out mean & 0.0481 & 0.0320 & 0.0254 \\
\midrule
Papers & 36 & 32 & 21 \\
Questions & 1459 & 1386 & 862 \\
\midrule
\multicolumn{4}{l}{\textit{Panel B. Sampling-variance-adjusted MSE (SE available)}} \\
\addlinespace[3pt]
\multicolumn{4}{l}{\textit{AI models}} \\
GPT-5.1 & 0.0247 & 0.0224 & 0.0163 \\
GPT-5.5 & 0.0163 & 0.0145 & 0.0098 \\
Opus 4.6 & 0.0185 & 0.0167 & 0.0099 \\
Opus 4.8 & --- & 0.0142 & 0.0101 \\
Fable 5 & --- & 0.0156 & 0.0092 \\
Astra & --- & --- & 0.0092 \\
\midrule
\multicolumn{4}{l}{\textit{Benchmarks}} \\
Always predict zero & 0.0524 & 0.0269 & 0.0214 \\
Leave-one-paper-out mean & 0.0483 & 0.0251 & 0.0211 \\
\midrule
Papers & 31 & 27 & 19 \\
Questions & 934 & 871 & 538 \\
\bottomrule\end{tabular}
\par\vspace{7pt}\begin{minipage}{0.97\linewidth}\footnotesize This table replicates Table 1, except applying equal weight to each question. Papers therefore are implicitly weighted in proportion to the number of questions they contain.\end{minipage}\end{table}

\clearpage
\begin{table}[p]\centering\small
\caption{Unadjusted forecast MSE on the SE-available sample}
\label{tab:accuracy-se-available}
\begin{tabular}{lrrr}
\toprule
 & Jan+ & Feb+ & May+ \\ \midrule
\multicolumn{4}{l}{\textit{AI models}} \\
GPT-5.1 & 0.0384 & 0.0380 & 0.0336 \\
GPT-5.5 & 0.0295 & 0.0297 & 0.0253 \\
Opus 4.6 & 0.0300 & 0.0302 & 0.0240 \\
Opus 4.8 & --- & 0.0282 & 0.0239 \\
Fable 5 & --- & 0.0294 & 0.0251 \\
Astra & --- & --- & 0.0236 \\
\midrule
\multicolumn{4}{l}{\textit{Benchmarks}} \\
Always predict zero & 0.0952 & 0.0570 & 0.0474 \\
Leave-one-paper-out mean & 0.0809 & 0.0469 & 0.0414 \\
\midrule
Papers & 31 & 27 & 19 \\
Questions & 934 & 871 & 538 \\
\bottomrule\end{tabular}
\par\vspace{7pt}\begin{minipage}{0.97\linewidth}\footnotesize This table reports unadjusted MSEs for the same sample and weights as Table~\ref{tab:accuracy}, Panel B. Standard errors are reported directly or recovered from reported test statistics and corresponding regression estimates, then expressed in the same units as the effects. \end{minipage}\end{table}

\clearpage
\begin{figure}[p]\centering
\includegraphics[width=\linewidth]{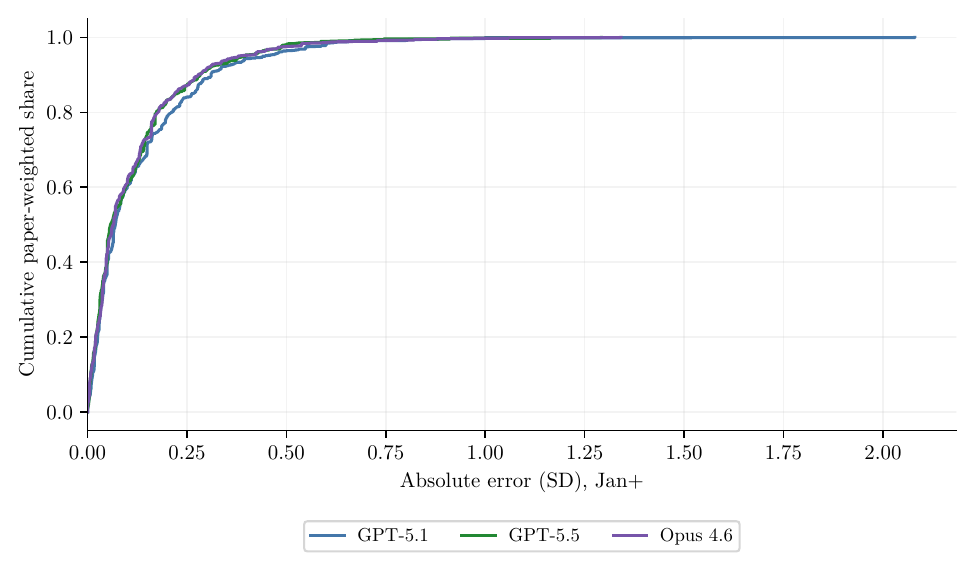}
\caption{Distribution of absolute forecast errors}
\label{fig:diagnostic-cdf}
\par
\begin{minipage}{0.97\linewidth}\footnotesize This figure plots the distribution of forecast errors across models for the time period with the most papers / questions, comparing GPT-5.1, GPT-5.5 and Opus 4.6 on all 1,459 Jan+ questions in 36 papers. Each question is equally weighted within a paper and each paper is equally weighted.\end{minipage}
\end{figure}

\clearpage
\begin{figure}[p]\centering
\includegraphics[width=\linewidth]{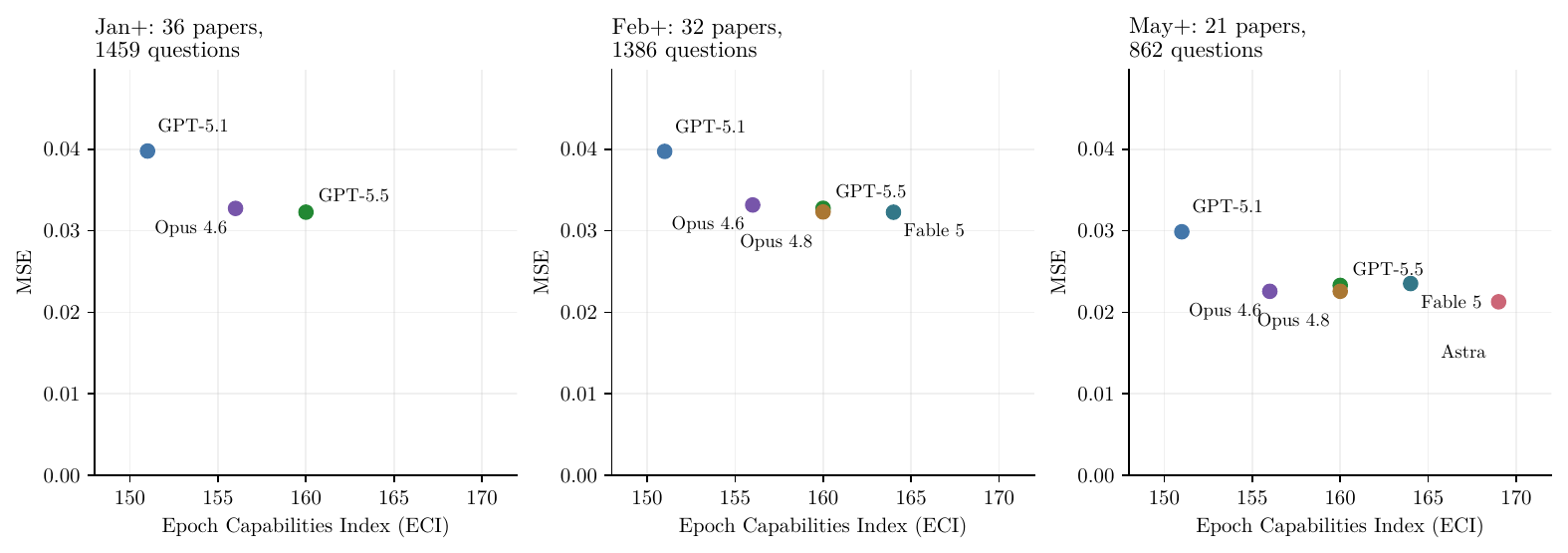}
\caption{Forecast MSE and general model capability}\label{fig:mse-eci}
\par
\begin{minipage}{0.97\linewidth}\footnotesize This figure shows the performance of different models against their ECI values. Within each panel, all models use identical questions; periods are nested. ECI values are from Epoch AI model pages, accessed November 14, 2026.\end{minipage}\end{figure}

\clearpage
\begin{landscape}
\begin{figure}[p]\centering
\includegraphics[width=\linewidth]{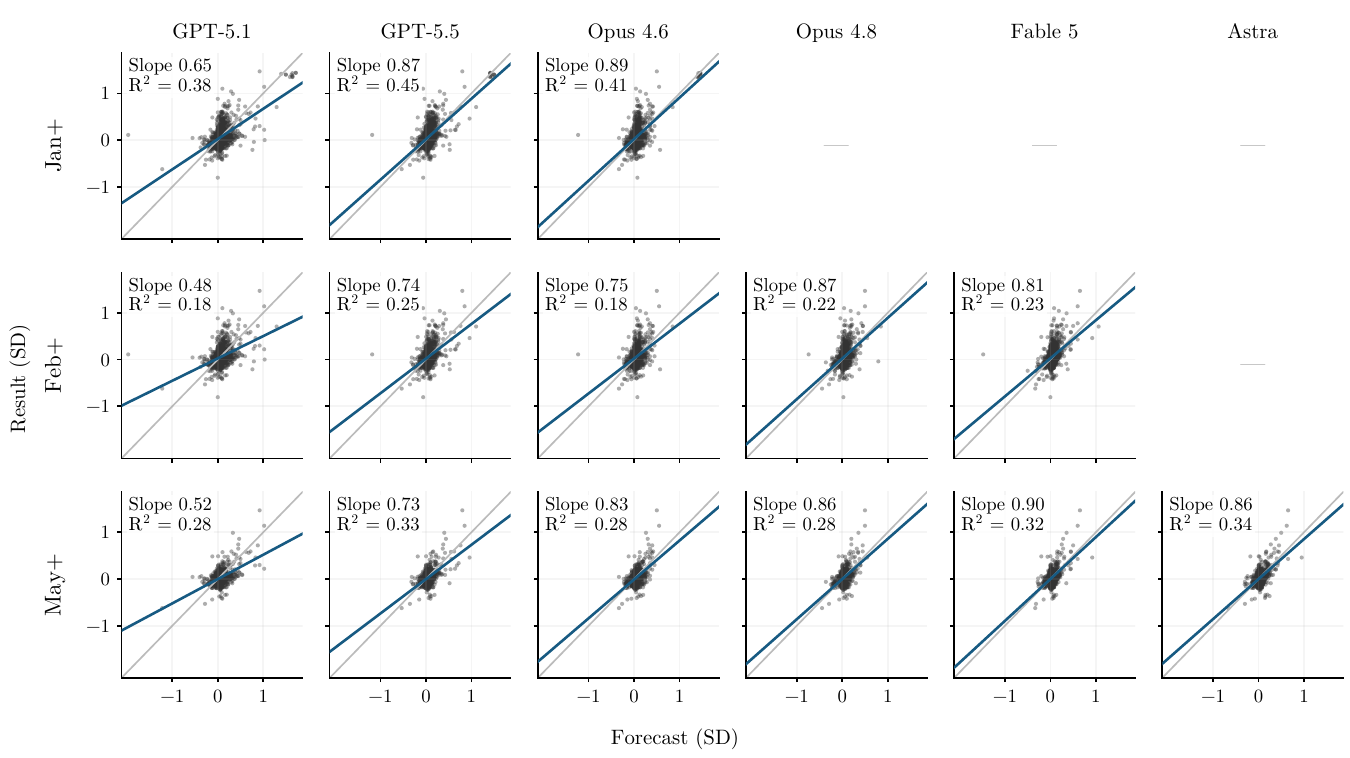}
\caption{Forecasts and reported treatment effects}
\label{fig:predictability-uncapped}
\par
\begin{minipage}{0.97\linewidth}\footnotesize
This figure shows scatterplots and lines of best fit between forecasts and results (both in standard deviations). The blue lines are fitted regressions; the gray diagonal is the 45-degree line where forecasts equal results. All observations are shown; slopes and $R^2$ use the full samples and weight questions equally. Rows show the Jan+ (1,459 questions, 36 papers), Feb+ (1,386 questions, 32 papers), and May+ (862 questions, 21 papers) samples, while columns show different models.
\end{minipage}
\end{figure}
\end{landscape}

\clearpage
\begin{figure}[p]\centering

\includegraphics[width=\linewidth]{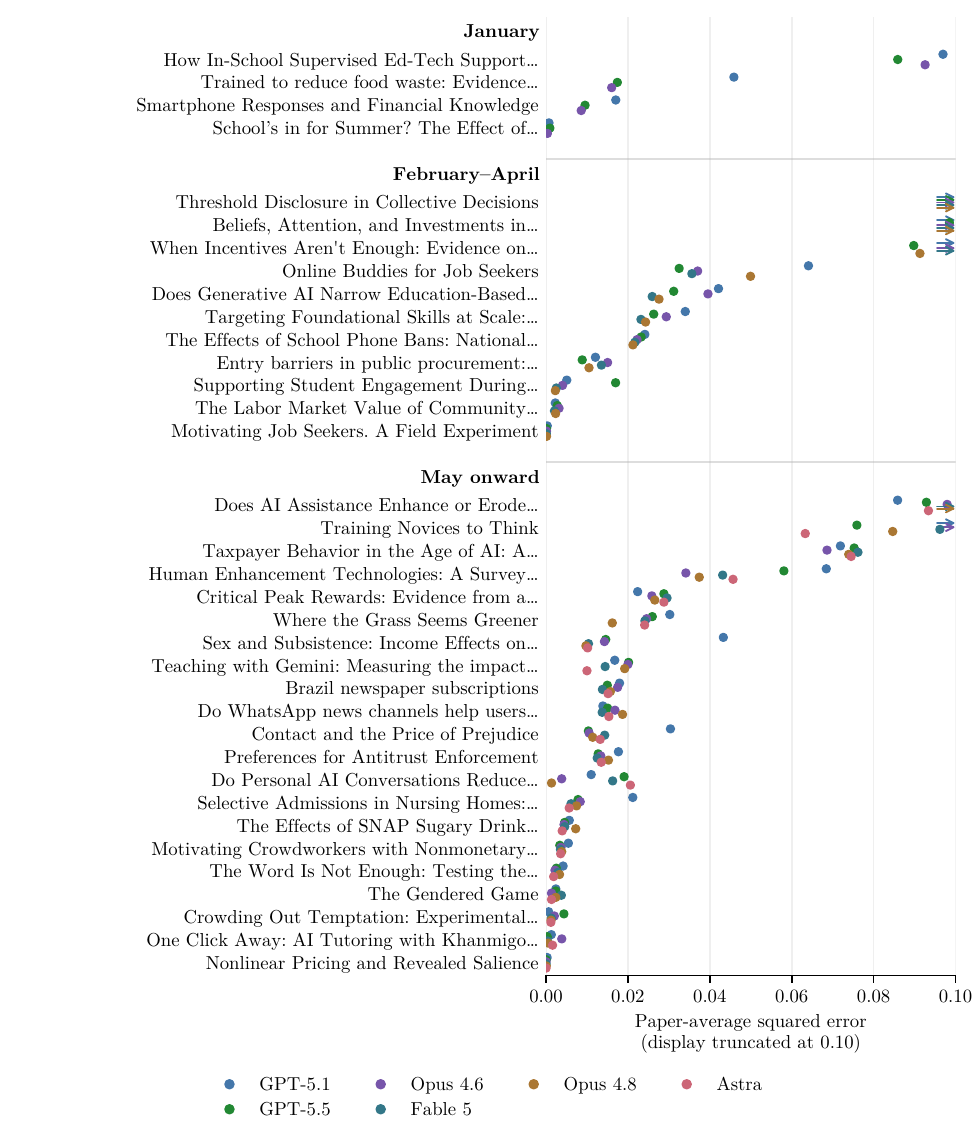}
\caption{Forecast accuracy by paper across the Jan+ sample}\label{fig:diagnostic-papers}
\par
\begin{minipage}{0.97\linewidth}\footnotesize This figure shows accuracy by paper and model. The sample is partitioned into nonoverlapping disclosure periods. Right arrows indicate the 16 paper--model MSEs above 0.10.
\end{minipage}\end{figure}

\clearpage
\begin{landscape}
\begin{figure}[p]\centering
\includegraphics[width=\linewidth]{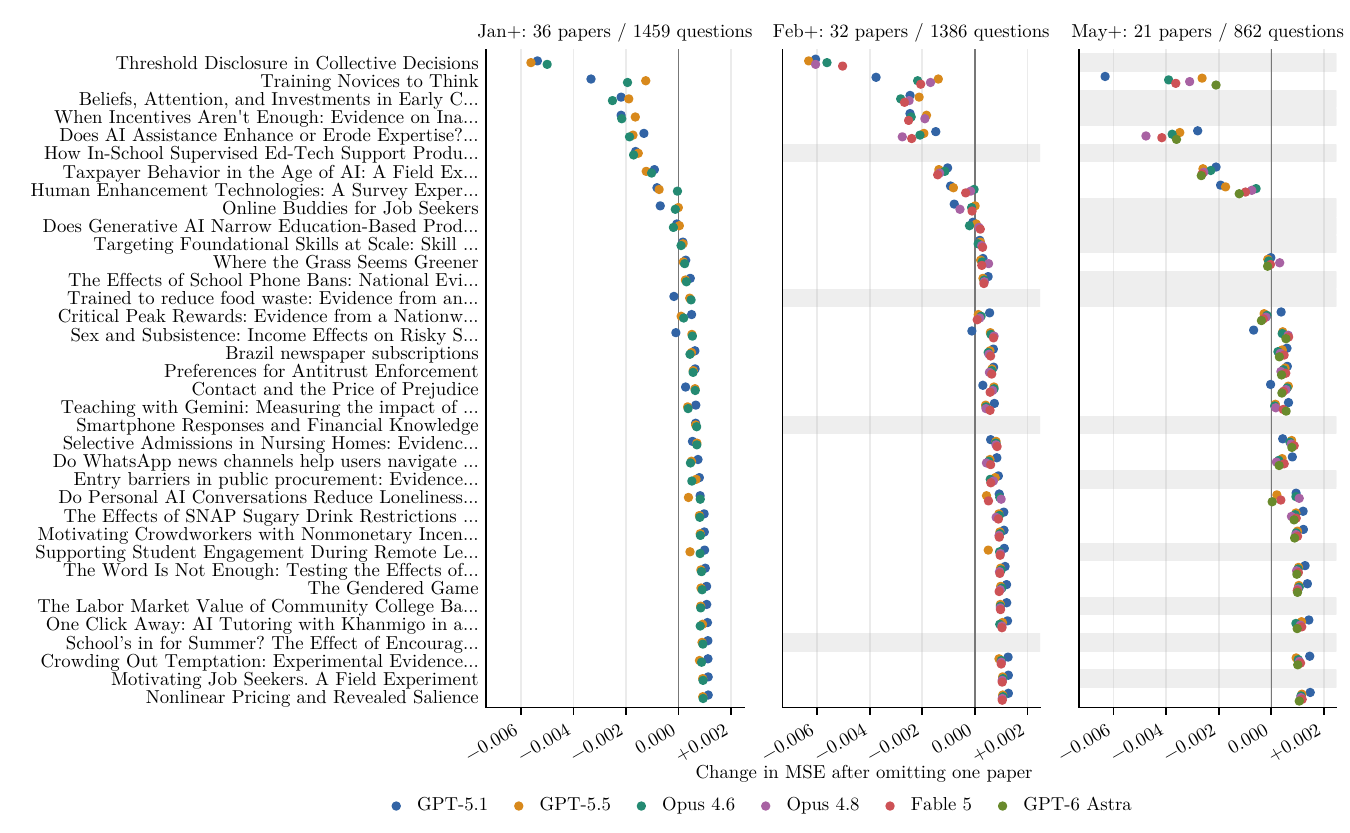}
\caption{Sensitivity to omitting one paper}\label{fig:paper-inclusion-combined}
\par
\begin{minipage}{0.97\linewidth}\footnotesize Each point is the change in the MSE after omitting a given paper. Negative values indicate lower MSE after omission. Shaded rows identify papers outside a panel's disclosure period.
\end{minipage}
\end{figure}
\end{landscape}

\clearpage
\begin{table}[p]\centering
\small
\caption{Repeatability of forecasts under identical requests}\label{tab:forecast-repeatability}
\setlength{\tabcolsep}{4pt}\begin{tabular}{llrrrrrr}\toprule
Model & Period & Papers & Questions & \shortstack{Mean absolute\\change (SD)} & Correlation & \shortstack{Original\\MSE} & \shortstack{Repeat\\MSE} \\ \midrule
GPT-5.1 & Jan+ & 36 & 1459 & 0.0398 & 0.9027 & 0.0398 & 0.0437 \\
GPT-5.5 & Jan+ & 36 & 1459 & 0.0187 & 0.9858 & 0.0323 & 0.0329 \\
Opus 4.6 & Jan+ & 36 & 1459 & 0.0135 & 0.9895 & 0.0327 & 0.0334 \\
Opus 4.8 & Feb+ & 32 & 1386 & 0.0155 & 0.9589 & 0.0323 & 0.0318 \\
Fable 5 & Feb+ & 32 & 1386 & 0.0153 & 0.9654 & 0.0323 & 0.0296 \\
Astra & May+ & 21 & 862 & 0.0106 & 0.9868 & 0.0213 & 0.0207 \\
\bottomrule\end{tabular}\par\vspace{7pt}\begin{minipage}{0.97\linewidth}\footnotesize This table shows how similar repeat runs of the same forecasting prompts were. The original run is the one used in the main analysis.
\end{minipage}\end{table}

\clearpage
\begin{figure}[p]\centering
\includegraphics[width=\linewidth]{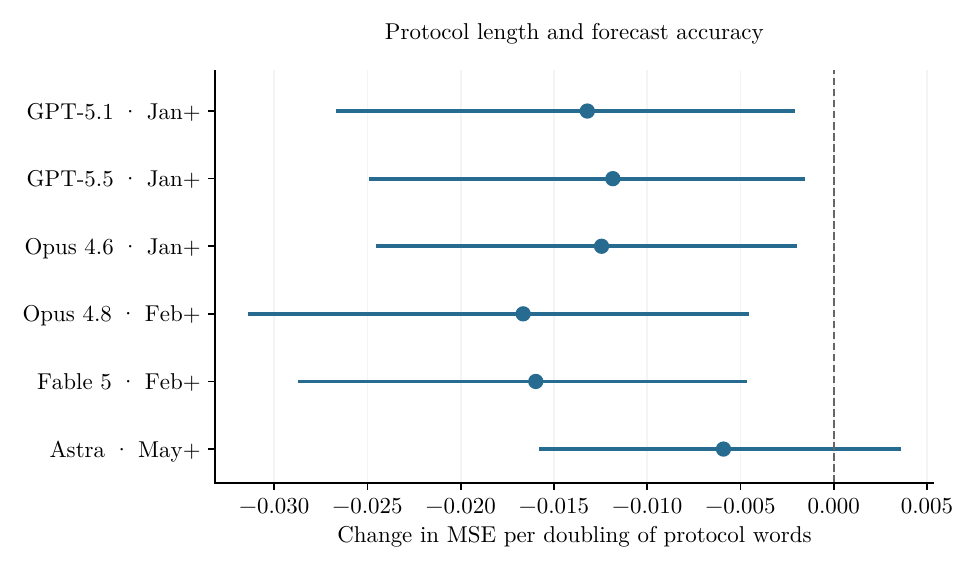}
\caption{Protocol length and forecast accuracy: regression coefficients}
\label{fig:protocol-length-coefficients}
\par
\begin{minipage}{0.97\linewidth}\footnotesize
This figure shows the estimated change in paper-level MSE per doubling of protocol word count, controlling for source and giving papers equal weight. Lines show 95\% percentile confidence intervals from 5,000 bootstrap resamples. Word counts use the protocol text supplied to the models, excluding instructions and forecasting questions.
\end{minipage}
\end{figure}

\clearpage
\begin{figure}[p]\centering
\includegraphics[width=\linewidth,height=0.75\textheight,keepaspectratio]{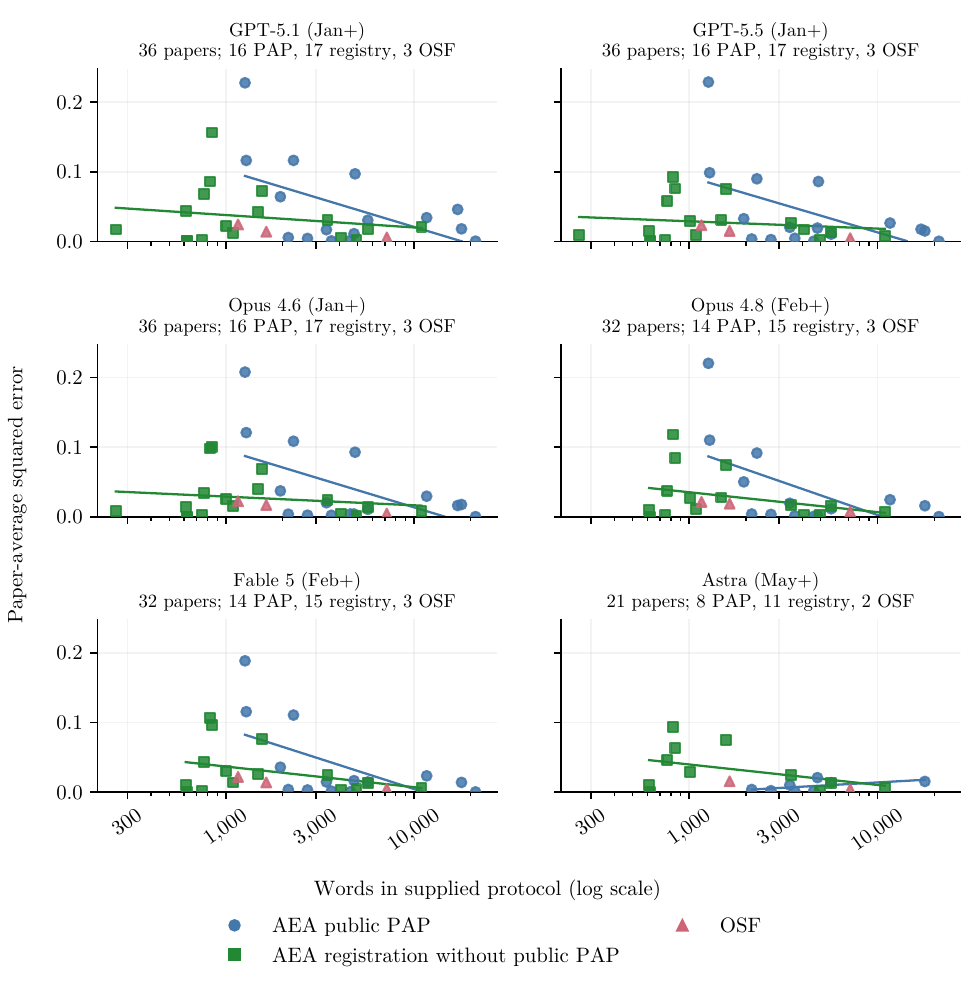}
\caption{Forecast accuracy and the length of supplied protocols}
\label{fig:protocol-length}
\par
\begin{minipage}{0.97\linewidth}\footnotesize
This figure shows detail in the relationship between protocol length and forecast accuracy. Each point is a paper. Solid lines fit paper-level MSE to log word count separately within each AEA source category, giving papers equal weight. No fitted line is shown for OSF because only two or three papers are available depending on the time period.
\end{minipage}
\end{figure}

\clearpage
\section{Sensitivity analyses}
\begin{figure}[!htbp]
\centering
\includegraphics[width=\textwidth]{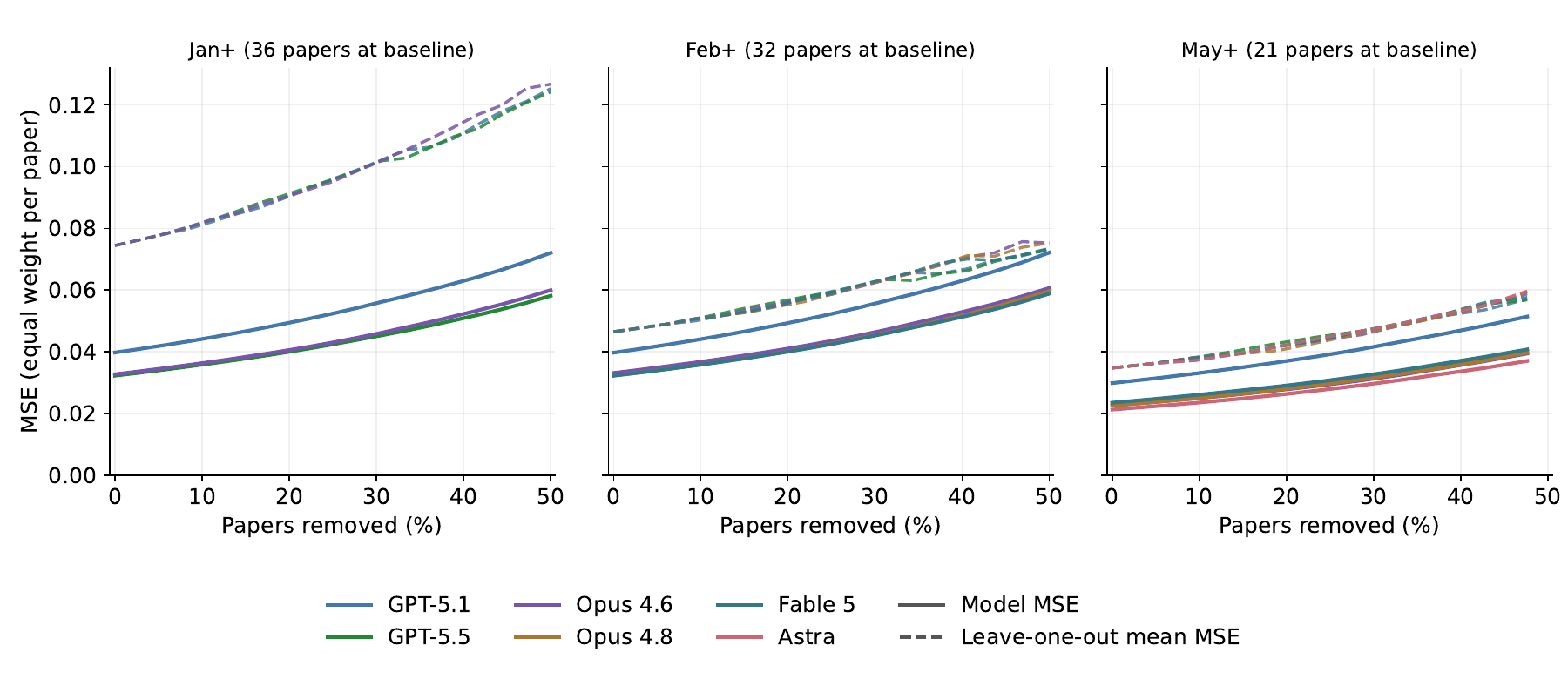}
\caption{Forecast accuracy after removing the most accurately forecast papers.}
\label{fig:paper-removal}
\par\begin{minipage}{0.97\linewidth}\footnotesize
Papers are removed in ascending order of their model-specific MSE, separately for each model and period. Solid curves show unadjusted model MSE; dashed curves of the same color show the leave-one-out mean benchmark recalculated on the same set of remaining papers. Different models can have different benchmarks since they can have different forecasting errors (i.e., the paper that is forecast most incorrectly by one model is not necessarily the same paper that is forecast the most incorrectly by another). Squared errors are averaged within paper and then equally weighted across papers. Each panel shows the removal of up to 50\% of the original sample: 18 of 36 papers for Jan+, 16 of 32 for Feb+, and 10 of 21 (47.6\%) for May+. %
\end{minipage}
\end{figure}

\clearpage
\subsection{Sensitivity to reasoning effort and output limits}
Table~\ref{tab:higher-effort} compares the highest-effort forecasts used in the main analysis with a high-effort alternative with a lower token cap. GPT-5.1 is excluded because high is already its maximum available effort. This alternative configuration increases the MSE for GPT-5.5 in May+ and Opus 4.8 in Feb+ and May+ and slightly decreases the MSE for Opus 4.6 in Jan+.

\begin{table}[p]\centering
\small
\caption{Sensitivity to reasoning effort and output limits}\label{tab:higher-effort}
\setlength{\tabcolsep}{4pt}\begin{tabular}{llrrr}\toprule
Model & Sample & \shortstack{Main\\MSE} & \shortstack{High-effort\\alternative MSE} & Change \\ \midrule
GPT-5.5 & Jan+ & 0.03228 & 0.03226 & -0.1\% \\
Opus 4.6 & Jan+ & 0.03274 & 0.03194 & -2.4\% \\
GPT-5.5 & Feb+ & 0.03276 & 0.03263 & -0.4\% \\
Opus 4.6 & Feb+ & 0.03316 & 0.03231 & -2.6\% \\
Opus 4.8 & Feb+ & 0.03231 & 0.03441 & +6.5\% \\
Fable 5 & Feb+ & 0.03227 & 0.03133 & -2.9\% \\
GPT-5.5 & May+ & 0.02330 & 0.02532 & +8.7\% \\
Opus 4.6 & May+ & 0.02257 & 0.02234 & -1.0\% \\
Opus 4.8 & May+ & 0.02258 & 0.02464 & +9.1\% \\
Fable 5 & May+ & 0.02353 & 0.02271 & -3.5\% \\
Astra & May+ & 0.02128 & 0.02185 & +2.6\% \\
\bottomrule\end{tabular}\par\vspace{7pt}\begin{minipage}{0.97\linewidth}\footnotesize The main specification uses xhigh for OpenAI and max for Anthropic, with a 64,000-token output cap, with the exception that GPT-5.1 is at high, its maximum available setting, and not shown here. The alternative uses high with a 32,000-token cap. Change is the percentage change in MSE relative to the main specification. Only the increases in the MSE for GPT-5.5 in May+ and Opus 4.8 in Feb+ and May+ and the decrease in the MSE for Opus 4.6 in Jan+ are significant at the 5\% level using 10,000 bootstrap draws of papers, with the same sampled papers used for both configurations.\end{minipage}\end{table}

\clearpage
\subsection{Sensitivity to forecasting prompts}\label{app:prompt-sensitivity}
The experiments in this section use the same model-specific effort settings and output limits as the main analysis.

\begin{figure}[p]
\centering

\includegraphics[width=\textwidth]{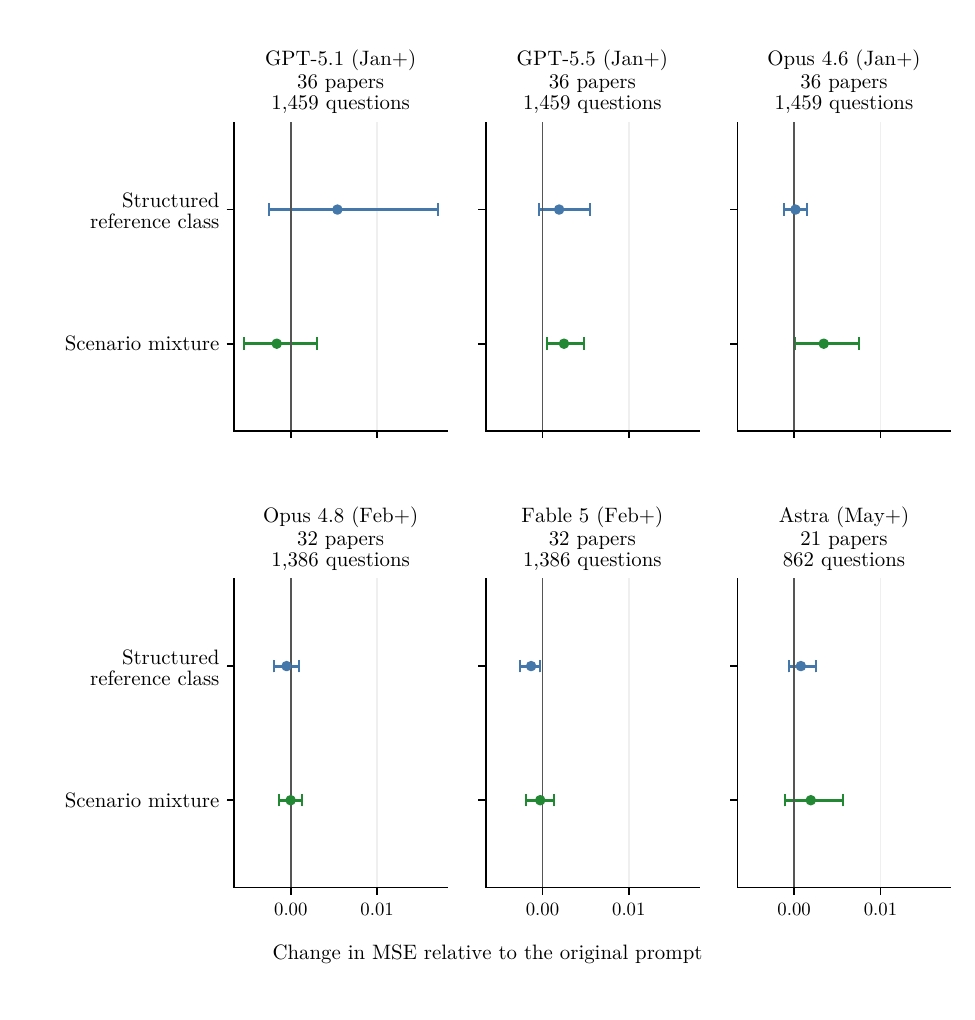}
\caption{Changes in forecast accuracy under alternative prompts.}
\label{fig:prompt-mse-changes-all-models}
\par
\begin{minipage}{0.97\linewidth}\footnotesize
The original prompt and both alternatives are compared on common valid questions within each model; papers receive equal weight. Negative values indicate lower MSE than the original prompt. 95\% percentile intervals from 10,000 paper bootstrap draws are shown, without multiplicity adjustment. One GPT-5.1 forecast with conflicting units and two missing Opus 4.6 forecasts were recovered through follow-up requests; each panel includes its full frozen analysis sample. Two Fable 5 forecasts explicitly expressed in percentage-point-style units are converted to shares before standardization.
\end{minipage}
\end{figure}

Figure~\ref{fig:prompt-mse-changes-all-models} compares two alternative prompts with the original prompt. One asks the LLM to consider relevant literature using a structured reference class; the other asks it to explicitly consider a mixture of scenarios. No prompt systematically improves accuracy across models.

\clearpage
\begin{table}[p]\centering\small
\caption{Sensitivity to source-based paper exclusions}
\label{tab:source-exclusions}
\setlength{\tabcolsep}{3pt}
\begin{tabular}{lrrrrrrrrcc}\toprule
\multicolumn{9}{c}{} & \multicolumn{2}{c}{Benchmarks} \\
\cmidrule(lr){10-11}
Exclusion & Papers & Questions & \shortstack{GPT-\\5.1} & \shortstack{GPT-\\5.5} & \shortstack{Opus\\4.6} & \shortstack{Opus\\4.8} & \shortstack{Fable\\5} & Astra & Zero & \shortstack{Other-\\paper\\mean} \\ \midrule
\multicolumn{11}{l}{\textit{Jan+}} \\ \addlinespace[3pt]
Main sample & 36 & 1,459 & 0.040 & 0.032 & 0.033 & -- & -- & -- & 0.086 & 0.074 \\
(A) & 34 & 1,380 & 0.041 & 0.033 & 0.034 & -- & -- & -- & 0.090 & 0.078 \\
(B) & 32 & 660 & 0.043 & 0.035 & 0.036 & -- & -- & -- & 0.095 & 0.081 \\
(C) & 35 & 1,445 & 0.041 & 0.033 & 0.034 & -- & -- & -- & 0.088 & 0.076 \\
(A), (B), and (C) & 29 & 567 & 0.046 & 0.038 & 0.038 & -- & -- & -- & 0.103 & 0.089 \\
\midrule
\multicolumn{11}{l}{\textit{Feb+}} \\ \addlinespace[3pt]
Main sample & 32 & 1,386 & 0.040 & 0.033 & 0.033 & 0.032 & 0.032 & -- & 0.054 & 0.047 \\
(A) & 30 & 1,307 & 0.041 & 0.034 & 0.034 & 0.034 & 0.034 & -- & 0.057 & 0.049 \\
(B) & 28 & 587 & 0.044 & 0.036 & 0.037 & 0.036 & 0.036 & -- & 0.060 & 0.051 \\
(C) & 31 & 1,372 & 0.041 & 0.034 & 0.034 & 0.033 & 0.033 & -- & 0.056 & 0.048 \\
(A), (B), and (C) & 25 & 494 & 0.047 & 0.039 & 0.040 & 0.039 & 0.039 & -- & 0.066 & 0.056 \\
\midrule
\multicolumn{11}{l}{\textit{May+}} \\ \addlinespace[3pt]
Main sample & 21 & 862 & 0.030 & 0.023 & 0.023 & 0.023 & 0.024 & 0.021 & 0.040 & 0.035 \\
(A) & 20 & 860 & 0.031 & 0.024 & 0.024 & 0.024 & 0.025 & 0.022 & 0.042 & 0.037 \\
(B) & 18 & 251 & 0.032 & 0.025 & 0.025 & 0.025 & 0.025 & 0.023 & 0.044 & 0.038 \\
(C) & 20 & 848 & 0.031 & 0.024 & 0.023 & 0.023 & 0.024 & 0.022 & 0.042 & 0.036 \\
(A), (B), and (C) & 16 & 235 & 0.036 & 0.028 & 0.027 & 0.028 & 0.028 & 0.025 & 0.049 & 0.042 \\
\bottomrule
\end{tabular}
\par\vspace{7pt}\begin{minipage}{0.97\linewidth}\footnotesize This table assesses sensitivity to various paper exclusion rules, as follows: (A) Exclude papers with main study statistics removed before forecasting. (B) Exclude papers with supplied pilot findings. (C) Exclude papers with hypothetical numerical results. The final row in each panel applies these exclusions together. %
\end{minipage}\end{table}

\clearpage
\section{Disclosure channels}\label{app:disclosure-channels}
Earlier findings were disclosed through several channels, sometimes before the paper itself became available. This section provides further detail. %

In one case, an experiment had multiple treatment arms whose results were reported in different papers, and a preprint existed for one treatment arm while results had not been reported for the second treatment arm.\footnote{This excludes ``Impact of a Long-Term Mentorship Program for Childbirth Providers on Quality of Care: A Cluster Randomized Controlled Trial'' from the main analysis.} Seminar abstracts and a conference abstract disclosed a small group of findings from three other experiments,\footnote{The seminar abstracts concerned ``Fun and Change: Video Game Edutainment Promotes Pro-Environmental Behaviour'' and ``Winning Is Contagious,'' the latter presented under the title ``Social Learning and the Dynamics of Union Support.'' The conference abstract concerned ``Does AI Help or Hurt Learning?''} while presentation slides and an earlier program report disclosed findings from two additional studies.\footnote{``Green Subsidies with Demand Distortions'' and ``If You Build It, They May Not Come: Willingness to Participate in Managed EV Charging,'' respectively.} Even when only a subset of results were publicly disclosed prior to the cut-off date, any paper with disclosed results was excluded from the main sample.

Media coverage provided another route through which results could enter the public domain. LLMs asked to screen the media for references that could pertain to a study were very comprehensive. For example, one paper was flagged for an author describing their findings on an outcome without specific reference to the experiment; in principle, the findings could have been in reference to a different study or program.\footnote{``Predictive Risk Scores in the Public Sector: Experimental Evidence from Child-Protection Investigations.''}

Any paper with potential disclosure of results was excluded from the analysis.

\clearpage
\section{Papers in the January-onward screened analysis sample}
\label{app:included-papers}
This list covers the 36 papers in the January-onward screened analysis sample. References identify the archived results papers; registry links identify their associated study records.
\begingroup\normalsize
\setlength{\parindent}{0pt}
\par\addvspace{0.7\baselineskip}\hangindent=1.5em\hangafter=1 Acemoglu, Daron, Nicolas Ajzenman, Guillermo Cruces, Martin Fiszbein, Gaston Garcia Zavaleta, and Carlos Molina. 2026. ``Where the Grass Seems Greener: Economic Misperceptions and Support for Democracy.'' NBER Working Paper 35644. \href{https://www.nber.org/papers/w35644}{Paper}. \href{https://www.socialscienceregistry.org/trials/12309}{AEARCTR-0012309}.\par
\par\addvspace{0.7\baselineskip}\hangindent=1.5em\hangafter=1 Acton, Riley K., Camila Morales, Julia A. Turner, Lois Miller, and Kalena Cortes. 2026. ``The Labor Market Value of Community College Bachelor's Degrees: Initial Evidence from a Resume Audit Study in Early Childhood Education.'' NBER Working Paper 35404. \href{https://www.nber.org/papers/w35404}{Paper}. \href{https://www.socialscienceregistry.org/trials/16106}{AEARCTR-0016106}.\par
\par\addvspace{0.7\baselineskip}\hangindent=1.5em\hangafter=1 Adena, Maja, Patrycja Janowska-Widomska, and Julian Harke. 2026. ``Motivating Crowdworkers with Nonmonetary Incentives and Payment Framing---Evidence from a Large-Scale Experiment.'' CRC TRR 190 Discussion Paper 576. \href{https://www.socialscienceregistry.org/trials/6744}{AEARCTR-0006744}.\par
\par\addvspace{0.7\baselineskip}\hangindent=1.5em\hangafter=1 Allcott, Hunt, Amy Finkelstein, Anna Grummon, and Matthew J. Notowidigdo. 2026. ``The Effects of SNAP Sugary Drink Restrictions on Consumption and Welfare.'' NBER Working Paper 35659. \href{https://www.nber.org/papers/w35659}{Paper}. \href{https://osf.io/5nf3b/}{OSF-5nf3b}.\par
\par\addvspace{0.7\baselineskip}\hangindent=1.5em\hangafter=1 Allcott, Hunt, E. Jason Baron, Thomas Dee, Angela L. Duckworth, Matthew Gentzkow, and Brian Jacob. 2026. ``The Effects of School Phone Bans: National Evidence from Lockable Pouches.'' NBER Working Paper 35132. \href{https://www.nber.org/papers/w35132}{Paper}. \href{https://osf.io/8twh9/}{OSF-8twh9}.\par
\par\addvspace{0.7\baselineskip}\hangindent=1.5em\hangafter=1 Andersen, Andreas Makoto Fukuda, Gwen-Jir\={o} Clochard, Guillaume Hollard, and Andreas Kotsadam. 2026. ``Contact and the Price of Prejudice.'' University of Osaka, ISER Discussion Paper 1317. \href{https://www.socialscienceregistry.org/trials/18089}{AEARCTR-0018089}.\par
\par\addvspace{0.7\baselineskip}\hangindent=1.5em\hangafter=1 Asirvatham, Hemanth, Chiara Betti, Rachel Brown, Arnaldo Camuffo, Aaron Chatterji, Chiara Fumagalli, Alfonso Gambardella, Myriam Mariani, Abhinav Pandey, Steve Ramos, Giovanni Salvucci, and Vladimir Simic. 2026. ``Training Novices to Think, or Giving Them LLMs? Evidence from an RCT.'' Working paper. \href{https://www.socialscienceregistry.org/trials/17124}{AEARCTR-0017124}.\par
\par\addvspace{0.7\baselineskip}\hangindent=1.5em\hangafter=1 Autor, David, Tanya Rodchenko, Josh Martin, Zanna Iscenko, Scott Strand, David Pearl, and Melissa Ferere. 2026. ``Does AI Assistance Enhance or Erode Expertise? Evidence from a Three-Month Field Experiment in Patent Drafting.'' NBER Working Paper 35720. \href{https://www.nber.org/papers/w35720}{Paper}. \href{https://www.socialscienceregistry.org/trials/15823}{AEARCTR-0015823}.\par
\par\addvspace{0.7\baselineskip}\hangindent=1.5em\hangafter=1 Bau, Natalie, Corinne Low, and Bryce Millett Steinberg. 2026. ``Sex and Subsistence: Income Effects on Risky Sexual Behavior and Partner Selection.'' NBER Working Paper 35940. \href{https://www.nber.org/papers/w35940}{Paper}. \href{https://www.socialscienceregistry.org/trials/14078}{AEARCTR-0014078}.\par
\par\addvspace{0.7\baselineskip}\hangindent=1.5em\hangafter=1 Bernard, Louise, Robert Metcalfe, and Andrew R. Schein. 2026. ``Critical Peak Rewards: Evidence from a Nationwide Demand Response Program.'' Working paper. \href{https://www.socialscienceregistry.org/trials/13034}{AEARCTR-0013034}.\par
\par\addvspace{0.7\baselineskip}\hangindent=1.5em\hangafter=1 Bhatt, Monica P., Jonathan Guryan, Fatemeh Momeni, Philip Oreopoulos, and Eleni Packis. 2026. ``Supporting Student Engagement During Remote Learning: Three Randomized Controlled Trials in Chicago Public Schools.'' NBER Working Paper 34996. \href{https://www.nber.org/papers/w34996}{Paper}. \href{https://www.socialscienceregistry.org/trials/6250}{AEARCTR-0006250}.\par
\par\addvspace{0.7\baselineskip}\hangindent=1.5em\hangafter=1 Braghieri, Luca, Leonardo Bursztyn, and Jan Fasnacht. 2026. ``Threshold Disclosure in Collective Decisions.'' Working paper. \href{https://www.socialscienceregistry.org/trials/16968}{AEARCTR-0016968}.\par
\par\addvspace{0.7\baselineskip}\hangindent=1.5em\hangafter=1 Burke, Jeremy, Carly Urban, and Olivia Valdes. 2026. ``Is Financial Knowledge Really Declining? Randomized Evidence on the Effects of Smartphone Responses.'' CESR-Schaeffer Working Paper 2026-004. \href{https://www.socialscienceregistry.org/trials/14002}{AEARCTR-0014002}.\par
\par\addvspace{0.7\baselineskip}\hangindent=1.5em\hangafter=1 Carrell, Scott E., Michal Kurlaender, Paco Martorell, and Christina Sun. 2026. ``School's in for Summer? The Effect of Encouraging Summer Community College Enrollment.'' AEA Papers and Proceedings 116: 362--366. \href{https://doi.org/10.1257/pandp.20261102}{Paper}. \href{https://www.socialscienceregistry.org/trials/11439}{AEARCTR-0011439}.\par
\par\addvspace{0.7\baselineskip}\hangindent=1.5em\hangafter=1 Castro, Caio and João F. Pugliese. 2026. ``Good News and Bad News: The Effects of Consuming Professional News Outlets During the 2022 Brazilian Elections.'' Working paper. \href{https://www.socialscienceregistry.org/trials/9940}{AEARCTR-0009940}.\par
\par\addvspace{0.7\baselineskip}\hangindent=1.5em\hangafter=1 Cockx, Bart, Johan Egebark, Greet Van Hoye, Emilie Videnord, and Johan Vikström. 2026. ``Motivating Job Seekers. A Field Experiment.'' IFAU Working Paper 2026:8. \href{https://www.socialscienceregistry.org/trials/5502}{AEARCTR-0005502}.\par
\par\addvspace{0.7\baselineskip}\hangindent=1.5em\hangafter=1 Cruces, Guillermo, Diego Fernández Meijide, Sebastian Galiani, Ramiro H. Gálvez, and María Lombardi. 2026. ``Does Generative AI Narrow Education-Based Productivity Gaps? Evidence from a Randomized Experiment.'' Universidad Torcuato Di Tella, Escuela de Gobierno, Documento de Trabajo 2026/03. \href{https://www.socialscienceregistry.org/trials/16607}{AEARCTR-0016607}.\par
\par\addvspace{0.7\baselineskip}\hangindent=1.5em\hangafter=1 Cunha, Flavio, Snejana Nihtianova, Jessica Rood, and Anja-Lize van der Merwe. 2026. ``Beliefs, Attention, and Investments in Early Childhood.'' NBER Working Paper 35150. \href{https://www.nber.org/papers/w35150}{Paper}. \href{https://www.socialscienceregistry.org/trials/9796}{AEARCTR-0009796}.\par
\par\addvspace{0.7\baselineskip}\hangindent=1.5em\hangafter=1 de Barros, Andreas and Theresa Lubozha. 2026. ``Targeting Foundational Skills at Scale: Skill Specificity and Transfer.'' CESifo Working Paper 12542. \href{https://www.socialscienceregistry.org/trials/14922}{AEARCTR-0014922}.\par
\par\addvspace{0.7\baselineskip}\hangindent=1.5em\hangafter=1 de Koning, Bart K., Paul Muller, Michèle Belot, Yvonne Engels, Didier Fouarge, Mario Keer, Philipp Kircher, and Sandra Phlippen. 2026. ``Online Buddies for Job Seekers: A Field Experiment.'' Working paper. \href{https://www.socialscienceregistry.org/trials/10322}{AEARCTR-0010322}.\par
\par\addvspace{0.7\baselineskip}\hangindent=1.5em\hangafter=1 Del Pozo Loayza, Cesar E., José D. López-Rivas, and Francisco Alpízar. 2026. ``Trained to reduce food waste: Evidence from an RCT with micro-sized and small enterprises in Cusco's hospitality industry.'' Journal of Environmental Economics and Management 138: 103355. \href{https://www.socialscienceregistry.org/trials/12063}{AEARCTR-0012063}.\par
\par\addvspace{0.7\baselineskip}\hangindent=1.5em\hangafter=1 Einav, Liran, Amy Finkelstein, Jesse Gubb, Neale Mahoney, and James Okun. 2026. ``Selective Admissions in Nursing Homes: Evidence from a Randomized Audit Study.'' NBER Working Paper 35944. \href{https://www.nber.org/papers/w35944}{Paper}. \href{https://www.socialscienceregistry.org/trials/18639}{AEARCTR-0018639}.\par
\par\addvspace{0.7\baselineskip}\hangindent=1.5em\hangafter=1 Fréget, Louis, Ariell Reshef, and Claudia Senik. 2026. ``Do Personal AI Conversations Reduce Loneliness? Evidence from a Pre-Registered Experiment.'' CESifo Working Paper 12882. \href{https://www.socialscienceregistry.org/trials/16749}{AEARCTR-0016749}.\par
\par\addvspace{0.7\baselineskip}\hangindent=1.5em\hangafter=1 Gennaioli, Nicola, Matthias Rodemeier, and Gregory Sun. 2026. ``Nonlinear Pricing and Revealed Salience.'' Working paper. \href{https://www.socialscienceregistry.org/trials/6581}{AEARCTR-0006581}.\par
\par\addvspace{0.7\baselineskip}\hangindent=1.5em\hangafter=1 Gorodnichenko, Yuriy, Ilona Sologoub, and Yuriy Fedyk. 2026. ``The Word Is Not Enough: Testing the Effects of Information Treatments on Perceived Corruption in Ukraine.'' NBER Working Paper 35251. \href{https://www.nber.org/papers/w35251}{Paper}. \href{https://www.socialscienceregistry.org/trials/13089}{AEARCTR-0013089}.\par
\par\addvspace{0.7\baselineskip}\hangindent=1.5em\hangafter=1 Holz, Justin E., Ricardo Perez-Truglia, Andrew Simon, and Alejandro Zentner. 2026. ``Taxpayer Behavior in the Age of AI: A Field Experiment on Property Tax Appeals.'' NBER Working Paper 35632. \href{https://www.nber.org/papers/w35632}{Paper}. \href{https://www.socialscienceregistry.org/trials/18430}{AEARCTR-0018430}.\par
\par\addvspace{0.7\baselineskip}\hangindent=1.5em\hangafter=1 Hyytinen, Ari, Jan Jääskeläinen, Antti Sieppi, Vesa-Heikki Soini, and Janne Tukiainen. 2026. ``Entry barriers in public procurement: Evidence from conjoint survey experiment.'' Aboa Centre for Economics Discussion Paper 176. \href{https://www.socialscienceregistry.org/trials/12628}{AEARCTR-0012628}.\par
\par\addvspace{0.7\baselineskip}\hangindent=1.5em\hangafter=1 Immordino, Giovanni, Mario Macis, Immacolata Marino, and Fabrizio Panebianco. 2026. ``Human Enhancement Technologies: A Survey Experiment on Private Demand and Governance Preferences.'' IZA Discussion Paper 18804. \href{https://www.socialscienceregistry.org/trials/16972}{AEARCTR-0016972}.\par
\par\addvspace{0.7\baselineskip}\hangindent=1.5em\hangafter=1 LearnLM Team, Google and Fab AI. 2026. ``Teaching with Gemini: Measuring the impact of Guided Learning on student mathematics progress in Sierra Leone.'' Research report, Google and Fab AI. \href{https://www.socialscienceregistry.org/trials/16651}{AEARCTR-0016651}.\par
\par\addvspace{0.7\baselineskip}\hangindent=1.5em\hangafter=1 Meinzen-Dick, Ruth, Thomas Falk, Hagar ElDidi, Richu Sanil, Katrina Kosec, and Lucia Carrillo. 2026. ``The Gendered Game: Effects of Experiential Learning on Decision-Making and Water Saving in India.'' IFPRI Discussion Paper 02436. \href{https://www.socialscienceregistry.org/trials/10765}{AEARCTR-0010765}.\par
\par\addvspace{0.7\baselineskip}\hangindent=1.5em\hangafter=1 Mont'Alverne, Camila, Patrícia Rossini, and Antonis Kalogeropoulos. 2026. ``Do WhatsApp news channels help users navigate the information environment? A field experiment on information-related attitudes.'' New Media \& Society, 1--21. \href{https://doi.org/10.1177/14614448261467150}{Paper}. \href{https://osf.io/fv945/}{OSF-fv945}.\par
\par\addvspace{0.7\baselineskip}\hangindent=1.5em\hangafter=1 Oreopoulos, Philip and Nina Low. 2026. ``One Click Away: AI Tutoring with Khanmigo in a Two-Year School Experiment.'' EdWorkingPaper 26-1551. \href{https://www.socialscienceregistry.org/trials/13519}{AEARCTR-0013519}.\par
\par\addvspace{0.7\baselineskip}\hangindent=1.5em\hangafter=1 Oreopoulos, Philip, Oliver Keyes-Krysakowski, and Deepak Agarwal. 2026. ``How In-School Supervised Ed-Tech Support Produces Massive Learning Gains: A Khan Academy Field Experiment in India.'' NBER Working Paper 34683. \href{https://www.nber.org/papers/w34683}{Paper}. \href{https://www.socialscienceregistry.org/trials/13921}{AEARCTR-0013921}.\par
\par\addvspace{0.7\baselineskip}\hangindent=1.5em\hangafter=1 Perez-Truglia, Ricardo, and Jeffrey Yusof. 2026. ``Who Wants to Break Up Big Firms? Harm, Fairness, and the Demand for Antitrust.'' NBER Working Paper 35503, revised August 2026. \href{https://www.nber.org/papers/w35503}{Paper}. \href{https://www.socialscienceregistry.org/trials/18249}{AEARCTR-0018249}.\par
\par\addvspace{0.7\baselineskip}\hangindent=1.5em\hangafter=1 Thirumurthy, Harsha, Elizabeth F. Bair, Dean Yang, Teniola I. Egbe, Benard Ayieko, Julius Wesonga, Sue Napierala, Connie Celum, Mary E. Putt, Carol S. Camlin, and Mark Ayallo. 2026. ``Crowding Out Temptation: Experimental Evidence on Savings and Health Behaviors in Kenya.'' NBER Working Paper 35833. \href{https://www.nber.org/papers/w35833}{Paper}. \href{https://www.socialscienceregistry.org/trials/9484}{AEARCTR-0009484}.\par
\par\addvspace{0.7\baselineskip}\hangindent=1.5em\hangafter=1 Yu, Hang, Jared Stolove, Dean Yang, James Riddell IV, and Arlete Mahumane. 2026. ``When Incentives Aren't Enough: Evidence on Inattention and Imperfect Memory from HIV Medication Adherence.'' NBER Working Paper 34917. \href{https://www.nber.org/papers/w34917}{Paper}. \href{https://www.socialscienceregistry.org/trials/3184}{AEARCTR-0003184}.\par
\endgroup

\clearpage
\section{Historical benchmark comparison}
\label{app:historical-benchmark}
\begin{figure}[p]\centering
\includegraphics[width=\linewidth]{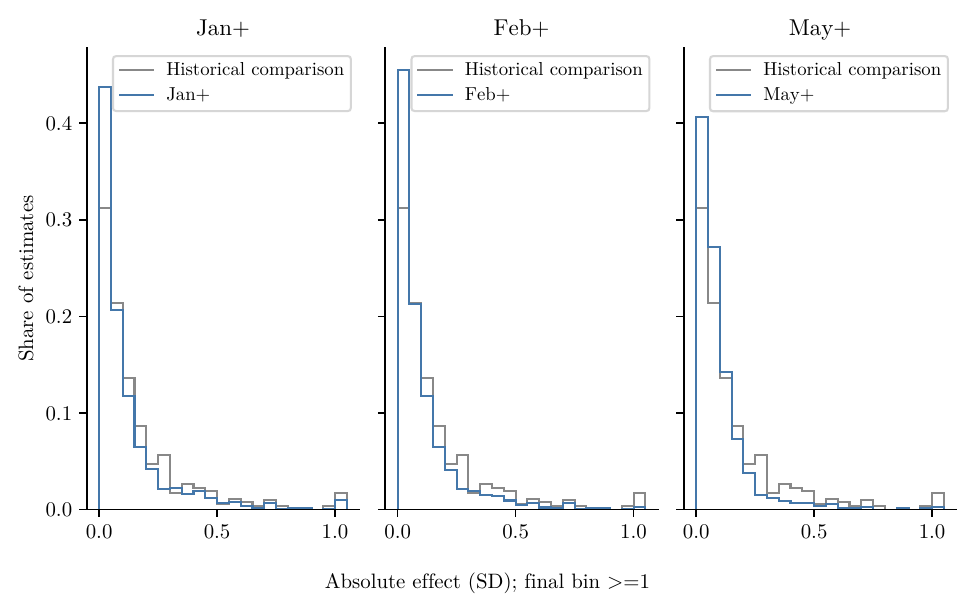}
\caption{Magnitudes of reported effects}
\label{fig:effect-histogram}
\par
\begin{minipage}{0.97\linewidth}\footnotesize This figure compares the set of papers available for AI forecasting in each time period with the historical results in \cite{vivalt2020}.\end{minipage}
\end{figure}

Results are shown in Table~\ref{tab:jeea-mse-tests}. Because the treatment effects in these data differ from those in the 2026 set I consider in this paper, they are not immediately comparable (Appendix Figure~\ref{fig:effect-histogram}). To make them more comparable, I rescale them so that their root-mean-square magnitude matches that of the Jan+, Feb+, and May+ groups. Samples still differ, but the leave-one-out meta-analyses based on \cite{vivalt2020} do not seem to predict treatment effects as well as LLMs. A researcher could in principle do better than a meta-analysis if they built a model of treatment effects for a particular intervention-outcome combination. For example, one might assume:
\begin{align*}
Y_i &= \theta_i + u_i, & u_i &\sim \mathcal{N}(0,\sigma_i^2), \\
\theta_i &= \alpha + X_i\beta + e_i, & e_i &\sim \mathcal{N}(0,\tau^2),
\end{align*}
where $Y_i$ is the estimated treatment effect in study $i$, $\theta_i$ is its true treatment effect, $X_i$ contains study characteristics, $\alpha$ is an intercept, and $\beta$ contains the corresponding coefficients. The sampling error $u_i$ has variance $\sigma_i^2$. $e_i$ is the residual study effect, with variance $\tau^2$ measuring residual heterogeneity remaining after accounting for $X_i$.

Table~\ref{tab:jeea-mse-tests} shows that closing the gap between the rescaled historical forecasts and the LLM forecasts would require reductions in $\tau^2$ of at least roughly 15--60\%, assuming prediction error consists of a component proportional to $\tau^2$ plus sampling noise. For context, in \cite{vivalt2020}, the single best-fitting covariate reduced estimated $\tau^2$ by about 10\% at the median and 20\% on average in the intervention-outcome combinations examined.

The main sample in Table~\ref{tab:jeea-mse-tests} used intervention-outcome combinations with at least three studies, but these may lead to noisy estimates. Table~\ref{tab:jeea-mse-tests-five} restricts the historical benchmark to intervention--outcome combinations with at least five studies. These appear to do slightly better, but not by much.

\clearpage
\begin{table}[p]\centering\small
\caption{AI MSE vs. MSE from historical benchmark}\label{tab:jeea-mse-tests}
\label{tab:break-even}
\setlength{\tabcolsep}{3pt}
\begin{tabular}{lrrrrrrr}\toprule
Model & \shortstack{AI\\MSE} & \shortstack{Original\\MSE} & $p$-value & \shortstack{Minimum $\tau^2$\\reduction\\required} & \shortstack{Rescaled\\MSE} & $p$-value & \shortstack{Minimum $\tau^2$\\reduction\\required} \\ \midrule
\multicolumn{8}{l}{Panel A. Jan+ (36 papers, 1459 questions)} \\
GPT-5.1 & 0.0398 & 0.0949 & 0.069 & 58\% & 0.0752 & 0.158 & 47\% \\
GPT-5.5 & 0.0323 & 0.0949 & 0.038 & 66\% & 0.0752 & 0.088 & 57\% \\
Opus 4.6 & 0.0327 & 0.0949 & 0.040 & 65\% & 0.0752 & 0.087 & 56\% \\
\midrule
\multicolumn{8}{l}{Panel B. Feb+ (32 papers, 1386 questions)} \\
GPT-5.1 & 0.0397 & 0.0949 & 0.072 & 58\% & 0.0475 & 0.282 & 16\% \\
GPT-5.5 & 0.0328 & 0.0949 & 0.041 & 65\% & 0.0475 & 0.042 & 31\% \\
Opus 4.6 & 0.0332 & 0.0949 & 0.042 & 65\% & 0.0475 & 0.031 & 30\% \\
Opus 4.8 & 0.0323 & 0.0949 & 0.040 & 66\% & 0.0475 & 0.022 & 32\% \\
Fable 5 & 0.0323 & 0.0949 & 0.039 & 66\% & 0.0475 & 0.026 & 32\% \\
\midrule
\multicolumn{8}{l}{Panel C. May+ (21 papers, 862 questions)} \\
GPT-5.1 & 0.0299 & 0.0949 & 0.033 & 69\% & 0.0352 & 0.577 & 15\% \\
GPT-5.5 & 0.0233 & 0.0949 & 0.019 & 75\% & 0.0352 & 0.181 & 34\% \\
Astra & 0.0213 & 0.0949 & 0.016 & 78\% & 0.0352 & 0.116 & 40\% \\
Opus 4.6 & 0.0226 & 0.0949 & 0.018 & 76\% & 0.0352 & 0.098 & 36\% \\
Opus 4.8 & 0.0226 & 0.0949 & 0.018 & 76\% & 0.0352 & 0.086 & 36\% \\
Fable 5 & 0.0235 & 0.0949 & 0.020 & 75\% & 0.0352 & 0.126 & 33\% \\
\bottomrule\end{tabular}
\par\vspace{7pt}\begin{minipage}{0.97\linewidth}\footnotesize
This table compares the MSE of the AI forecasts using the AEA RCT Registry and OSF with the MSE one would get by using leave-one-out meta-analysis to predict the left out study across the 47 intervention-outcome combinations with at least 3 papers in \cite{vivalt2020}. These samples are different, but perhaps still informative. Predictions use random-effects inverse-variance weights $1/(SE_i^2+\widehat{\tau}^2)$, with $\widehat{\tau}^2$ estimated by restricted maximum likelihood using the remaining studies. The ``Original MSE'' column builds off the original estimates while the ``Rescaled MSE'' column multiplies the historical effects and leave-one-out predictions by the ratio of the AI sample's root-mean-square reported effect to its historical counterpart, separately for each panel, to partially adjust for the different samples. The $p$-values test equality of the AI MSE compared to the MSE implied by the data in \cite{vivalt2020}. Bootstrap tests use 50,000 draws independently resampling AI papers and historical groups. Rescaling is recomputed in every draw. The minimum required reduction in $\tau^2$ is $\max\{0,1-\mathrm{MSE}_{\mathrm{AI}}/\mathrm{MSE}_{\mathrm{H}}\}$, using the original or rescaled historical MSE. %
\end{minipage}\end{table}

\clearpage
\begin{table}[p]\centering\small
\caption{AI MSE vs. MSE from historical benchmark: intervention-outcomes with at least 5 papers}\label{tab:jeea-mse-tests-five}
\setlength{\tabcolsep}{3pt}
\begin{tabular}{lrrrrrrr}\toprule
Model & \shortstack{AI\\MSE} & \shortstack{Original\\MSE} & $p$-value & \shortstack{Minimum $\tau^2$\\reduction\\required} & \shortstack{Rescaled\\MSE} & $p$-value & \shortstack{Minimum $\tau^2$\\reduction\\required} \\ \midrule
\multicolumn{8}{l}{Panel A. Jan+ (36 papers, 1459 questions)} \\
GPT-5.1 & 0.0398 & 0.0795 & 0.145 & 50\% & 0.0745 & 0.173 & 47\% \\
GPT-5.5 & 0.0323 & 0.0795 & 0.077 & 59\% & 0.0745 & 0.087 & 57\% \\
Opus 4.6 & 0.0327 & 0.0795 & 0.082 & 59\% & 0.0745 & 0.085 & 56\% \\
\midrule
\multicolumn{8}{l}{Panel B. Feb+ (32 papers, 1386 questions)} \\
GPT-5.1 & 0.0397 & 0.0795 & 0.148 & 50\% & 0.0471 & 0.321 & 16\% \\
GPT-5.5 & 0.0328 & 0.0795 & 0.083 & 59\% & 0.0471 & 0.050 & 30\% \\
Opus 4.6 & 0.0332 & 0.0795 & 0.086 & 58\% & 0.0471 & 0.036 & 30\% \\
Opus 4.8 & 0.0323 & 0.0795 & 0.081 & 59\% & 0.0471 & 0.024 & 31\% \\
Fable 5 & 0.0323 & 0.0795 & 0.078 & 59\% & 0.0471 & 0.030 & 31\% \\
\midrule
\multicolumn{8}{l}{Panel C. May+ (21 papers, 862 questions)} \\
GPT-5.1 & 0.0299 & 0.0795 & 0.066 & 62\% & 0.0349 & 0.601 & 14\% \\
GPT-5.5 & 0.0233 & 0.0795 & 0.034 & 71\% & 0.0349 & 0.194 & 33\% \\
Astra & 0.0213 & 0.0795 & 0.029 & 73\% & 0.0349 & 0.122 & 39\% \\
Opus 4.6 & 0.0226 & 0.0795 & 0.033 & 72\% & 0.0349 & 0.106 & 35\% \\
Opus 4.8 & 0.0226 & 0.0795 & 0.034 & 72\% & 0.0349 & 0.091 & 35\% \\
Fable 5 & 0.0235 & 0.0795 & 0.036 & 70\% & 0.0349 & 0.137 & 33\% \\
\bottomrule\end{tabular}
\par\vspace{7pt}\begin{minipage}{0.97\linewidth}\footnotesize
This table compares the MSE of the AI forecasts using the AEA RCT Registry and OSF with the MSE one would get by using leave-one-out meta-analysis to predict the left out study across the 34 intervention-outcome combinations with at least 5 papers in \cite{vivalt2020}. These samples are different, but perhaps still informative. Predictions use random-effects inverse-variance weights $1/(SE_i^2+\widehat{\tau}^2)$, with $\widehat{\tau}^2$ estimated by restricted maximum likelihood using the remaining studies. The ``Original MSE'' column builds off the original estimates while the ``Rescaled MSE'' column multiplies the historical effects and leave-one-out predictions by the ratio of the AI sample's root-mean-square reported effect to its historical counterpart, separately for each panel, to partially adjust for the different samples. The $p$-values test equality of the AI MSE compared to the MSE implied by the data in \cite{vivalt2020}. Bootstrap tests use 50,000 draws independently resampling AI papers and historical groups. Rescaling is recomputed in every draw. The minimum required reduction in $\tau^2$ is $\max\{0,1-\mathrm{MSE}_{\mathrm{AI}}/\mathrm{MSE}_{\mathrm{H}}\}$, using the original or rescaled historical MSE. %
\end{minipage}\end{table}

\clearpage
\section{Online AI appendix}

Table~\ref{tab:protocol-summary} provides further detail on which papers were excluded using each of the rules in Table~\ref{tab:source-exclusions}. Table~\ref{tab:protocol-information} summarizes the relevant dates and number of questions included from each paper in each time period. Both tables were compiled using Astra.

\clearpage
\begin{table}[p]\centering\footnotesize\setstretch{1}
\caption{Summary of information in forecasting protocols}
\label{tab:protocol-summary}
\begin{tabular}{p{0.23\linewidth}p{0.23\linewidth}p{0.44\linewidth}}\toprule
Sensitivity-analysis exclusion rule & Studies & Information in source materials \\ \midrule
(A) Exclude papers with main study statistics removed before forecasting & 2 studies: Khanmigo; Targeting Foundational Skills at Scale & Khanmigo: tables reporting actual student counts in the analysis samples, time spent using the platform, platform activity, and first-stage estimates were removed from the source text. Targeted Instruction: uptake and control-group endline information were removed. The forecasters received the remaining methods and definitions. \\ \addlinespace[8pt]
(B) Exclude papers with supplied pilot findings & 5 studies: Motivating Job Seekers; Brazil newspaper subscriptions; Informed Choices; AI Conversations and Loneliness; Selective Admissions in Nursing Homes & The supplied plans describe observed pilot responses, usage, measurement properties or outcome levels. Nursing homes includes racial differences of 15, 11 and 20 percentage points across three response outcomes; Brazil includes coupon redemption and news-use figures. Informed Choices contributes no questions to the screened sample. \\ \addlinespace[8pt]
(C) Exclude papers with hypothetical numerical results & 1 study: SNAP Sugary Drink Restrictions & Example tables contain numerical coefficients labeled as hypothetical illustrations by the authors. \\ \addlinespace[8pt]
\bottomrule\end{tabular}
\par\vspace{7pt}\begin{minipage}{0.97\linewidth}\footnotesize Categories (A), (B), and (C) match Table~\ref{tab:source-exclusions}. Counts here refer to the 39 audited studies; Table~\ref{tab:source-exclusions} reports the papers and questions remaining in each analysis sample after exclusion. Informed Choices contributes no screened questions. Categories can overlap.\end{minipage}\end{table}

\clearpage
\begingroup\footnotesize\setstretch{1}
\setlength{\tabcolsep}{4pt}
\begin{longtable}{p{0.29\linewidth}p{0.25\linewidth}p{0.405\linewidth}}
\caption{Paper-by-paper information in forecasting protocols}\label{tab:protocol-information}\\
\toprule Study; questions (Jan+ / Feb+ / May+) & Document version and timing & Information supplied to forecasters \\ \midrule
\endfirsthead
\multicolumn{3}{l}{\textit{Table \thetable\ continued}}\\
\toprule Study; questions (Jan+ / Feb+ / May+) & Document version and timing & Information supplied to forecasters \\ \midrule
\endhead
\midrule\multicolumn{3}{r}{\textit{Continued on next page}}\\\endfoot
\bottomrule\endlastfoot
When Incentives Aren't Enough: Evidence on Inattention and Imperfect Memory from HIV Medication Adherence\newline AEARCTR-0003184\newline 99 / 99 / 0 & Registry first-registered field: 10 Sep 2018. Recorded registry version 5.0 published: 21 Jan 2026. Plan dated 14 Aug 2019. & Intervention, outcomes and planned analysis are described. An earlier population survey informs the stigma intervention. \\ \addlinespace[7pt]
Motivating Job Seekers. A Field Experiment\newline AEARCTR-0005502\newline 188 / 188 / 0 & Registry first-registered field: 26 Feb 2020. Recorded registry version 3.1 published: 23 Jul 2025. Plan dated 20 Sep 2022 (update). & Pilot response/manipulation findings and historical outcome moments; power calculations also use early-cohort information pooled across treatment arms. \\ \addlinespace[7pt]
Supporting Student Engagement During Remote Learning: Three Randomized Controlled Trials in Chicago Public Schools\newline AEARCTR-0006250\newline 15 / 15 / 0 & Registry first-registered field: 17 Sep 2020. Recorded registry version 1.0 published: 18 Sep 2020. & Realized treatment/control allocation counts and procedural descriptions. \\ \addlinespace[7pt]
Nonlinear Pricing and Revealed Salience\newline AEARCTR-0006581\newline 1 / 1 / 1 & Registry first-registered field: 11 Oct 2020. Recorded registry version 1.2 published: 22 Jun 2026. & Intervention, outcomes and planned analysis are described. \\ \addlinespace[7pt]
Motivating Crowdworkers with Nonmonetary Incentives and Payment Framing—Evidence from a Large-Scale Experiment\newline AEARCTR-0006744\newline 55 / 55 / 55 & Registry first-registered field: 1 Dec 2020. Recorded registry version 1.3 published: 6 Jun 2024. Plan dated 14 Jan 2021. & Planned payment-framing/recognition experiment and recruiting-platform population counts. \\ \addlinespace[7pt]
Crowding Out Temptation: Experimental Evidence on Savings and Health Behaviors in Kenya\newline AEARCTR-0009484\newline 6 / 6 / 6 & Registry first-registered field: 20 May 2024. Recorded registry version 2.0 published: 15 Apr 2025. Plan dated 7 Apr 2025 (version 1). & References a previous pilot to describe intervention design; states the plan was finalized after baseline collection and before investigator unblinding. \\ \addlinespace[7pt]
Beliefs, Attention, and Investments in Early Childhood\newline AEARCTR-0009796\newline 5 / 5 / 0 & Registry first-registered field: 11 Oct 2022. Recorded registry version 1.0 published: 17 Oct 2022. & Intervention, outcomes and planned analysis are described. \\ \addlinespace[7pt]
Brazil newspaper subscriptions\newline AEARCTR-0009940\newline 319 / 319 / 319 & Registry first-registered field: 1 Sep 2022. Recorded registry version 1.2 published: 26 Jun 2023. Plan dated 26 Nov 2022. & Numerical pilot coupon-redemption and news-use findings, pilot sample composition and power inputs are supplied. \\ \addlinespace[7pt]
Online Buddies for Job Seekers\newline AEARCTR-0010322\newline 47 / 47 / 0 & Registry first-registered field: 27 Oct 2022. Recorded registry version 1.1 published: 9 Jan 2026. Plan dated 27 Oct 2022. & Recruitment-to-date counts and assumed control means in power calculations. \\ \addlinespace[7pt]
The Gendered Game\newline AEARCTR-0010765\newline 1 / 1 / 1 & Registry first-registered field: 13 Jan 2023. Recorded registry version 1.0 published: 23 Jan 2023. & Documents completed baseline and intervention rounds and earlier survey covariates used in sampling. \\ \addlinespace[7pt]
School’s in for Summer? The Effect of Encouraging Summer Community College Enrollment\newline AEARCTR-0011439\newline 2 / 0 / 0 & Registry first-registered field: 16 May 2023. Recorded registry version 1.0 published: 19 May 2023. & Intervention, outcomes and planned analysis are described. \\ \addlinespace[7pt]
Trained to reduce food waste: Evidence from an RCT with micro-sized and small enterprises in Cusco's hospitality industry\newline AEARCTR-0012063\newline 18 / 0 / 0 & Registry first-registered field: 6 Nov 2023. Recorded registry version 2.0 published: 1 Dec 2023. Plan dated 11 Nov 2023. & Numerical pre-intervention firm-census characteristics, correlations and stratification coefficients; planned pilot audit and power calculations. \\ \addlinespace[7pt]
Where the Grass Seems Greener\newline AEARCTR-0012309\newline 14 / 14 / 14 & Registry first-registered field: 16 Oct 2023. Recorded registry version 1.1 published: 18 Oct 2023. & Intervention, outcomes and planned analysis are described. Contains alternative design/sample descriptions. \\ \addlinespace[7pt]
Entry barriers in public procurement: Evidence from conjoint survey experiment\newline AEARCTR-0012628\newline 20 / 20 / 0 & Registry first-registered field: 29 Nov 2023. Recorded registry version 1.0 published: 6 Dec 2023. & Intervention, outcomes and planned analysis are described. Numeric vignette attributes are experimental treatments. \\ \addlinespace[7pt]
Critical Peak Rewards: Evidence from a Nationwide Demand Response Program\newline AEARCTR-0013034\newline 1 / 1 / 1 & Registry first-registered field: 13 Feb 2024. Recorded registry version 1.0 published: 16 Feb 2024. & Intervention emails, incentive amounts, and randomization details. \\ \addlinespace[7pt]
The Word Is Not Enough: Testing the Effects of Information Treatments on Perceived Corruption in Ukraine\newline AEARCTR-0013089\newline 1 / 1 / 1 & Registry first-registered field: 27 Feb 2024. Recorded registry version 1.0 published: 6 Mar 2024. & Describes a January 2024 pilot with 49 respondents. \\ \addlinespace[7pt]
One Click Away: AI Tutoring with Khanmigo in a Two-Year School Experiment\newline AEARCTR-0013519\newline 2 / 2 / 2 & Registry first-registered field: 28 Apr 2024. Recorded registry version 2.0 published: 14 May 2026. Plan dated 13 May 2026. Earlier experimental-details posting: 9 May 2024. & May 2026 analysis plan containing observed sample, balance, and platform-use statistics. Filled diagnostic and first-stage tables were removed before both question generation and forecasting. The plan describes the analysis hierarchy as fixed before analysis. \\ \addlinespace[7pt]
Agrivoltaics can reduce political polarization and local opposition to solar energy on land\newline AEARCTR-0013540\newline 0 / 0 / 0 & Registry first-registered field: 1 May 2024. Recorded registry version 1.1 published: 28 Oct 2024. & Intervention, outcomes and planned analysis are described. \\ \addlinespace[7pt]
How In-School Supervised Ed-Tech Support Produces Massive Learning Gains: A Khan Academy Field Experiment in India\newline AEARCTR-0013921\newline 47 / 0 / 0 & Registry first-registered field: 1 Jul 2024. Recorded registry version 2.0 published: 3 Jul 2025. Supplied plan is undated. & The supplied text contains observed sample counts, exclusions, and baseline-balance statistics. The text also includes endline-effect and heterogeneity table shells, figure references, and captions. \\ \addlinespace[7pt]
Smartphone Responses and Financial Knowledge\newline AEARCTR-0014002\newline 6 / 0 / 0 & Registry first-registered field: 19 Jul 2024. Recorded registry version 1.0 published: 29 Jul 2024. & References prior descriptive evidence motivating survey-mode randomization. \\ \addlinespace[7pt]
Incentivizing Self-Protection From Wildfires\newline AEARCTR-0014003\newline 0 / 0 / 0 & Registry first-registered field: 10 Jul 2024. Recorded registry version 1.0 published: 16 Jul 2024. & Describes an earlier wildfire pilot. \\ \addlinespace[7pt]
Sex and Subsistence: Income Effects on Risky Sexual Behavior and Partner Selection\newline AEARCTR-0014078\newline 3 / 3 / 3 & Registry first-registered field: 12 Aug 2024. Recorded registry version 1.0 published: 14 Aug 2024. & Qualitative background on vulnerability and risky coping. \\ \addlinespace[7pt]
Targeting Foundational Skills at Scale: Skill Specificity and Transfer\newline AEARCTR-0014922\newline 77 / 77 / 0 & Registry first-registered field: 27 Nov 2024. Recorded registry version 1.5 published: 15 Apr 2025. Plan dated 14 Apr 2025. & Baseline/descriptive material and documented analysis amendments. The amendments describe measurement and covariate decisions using control-group follow-up data. Main study uptake and control-group endline figures were removed before submission. \\ \addlinespace[7pt]
Does AI Assistance Enhance or Erode Expertise? Evidence from a Three-Month Field Experiment in Patent Drafting\newline AEARCTR-0015823\newline 32 / 32 / 32 & Registry first-registered field: 12 May 2025. Recorded registry version 1.0 published: 21 May 2025. & Intervention, outcomes and planned analysis are described. Baseline patent drafts are part of planned measurement. \\ \addlinespace[7pt]
Informed Choices: Gender Gaps in Career Advice\newline AEARCTR-0015913\newline 0 / 0 / 0 & Registry first-registered field: 28 Apr 2025. Recorded registry version 1.1 published: 12 Sep 2025. & Pilot control means for expected work hours and information-session interest appear in power calculations. \\ \addlinespace[7pt]
The Labor Market Value of Community College Bachelor’s Degrees: Initial Evidence from a Resume Audit Study in Early Childhood Education\newline AEARCTR-0016106\newline 7 / 7 / 0 & Registry first-registered field: 29 May 2025. Recorded registry version 1.1 published: 22 May 2026. & Describes the design of a planned pilot resume-audit study. \\ \addlinespace[7pt]
Does Generative AI Narrow Education-Based Productivity Gaps? Evidence from a Randomized Experiment\newline AEARCTR-0016607\newline 16 / 16 / 0 & Registry first-registered field: 28 Aug 2025. Recorded registry version 1.1 published: 14 Feb 2026. & Intervention, outcomes and planned analysis are described. \\ \addlinespace[7pt]
Teaching with Gemini: Measuring the impact of Guided Learning on student mathematics progress in Sierra Leone\newline AEARCTR-0016651\newline 15 / 15 / 15 & Registry first-registered field: 18 Sep 2025. Recorded registry version 2.2 published: 14 Jul 2026. Supplied plan is undated. & Analysis commitments and measurement definitions. \\ \addlinespace[7pt]
Do Personal AI Conversations Reduce Loneliness? Evidence from a Pre-Registered Experiment\newline AEARCTR-0016749\newline 17 / 17 / 17 & Registry first-registered field: 3 Jan 2026. Recorded registry version 2.0 published: 17 Jul 2026. Plan dated 3 Jan 2026. & Pilot of 1,000 participants: no differential attrition, scale reliability, and baseline loneliness median inform design and measurement. \\ \addlinespace[7pt]
Threshold Disclosure in Collective Decisions\newline AEARCTR-0016968\newline 15 / 15 / 0 & Registry first-registered field: 8 Oct 2025. Recorded registry version 1.0 published: 22 Oct 2025. Plan dated Oct 2025. & Intervention, outcomes and planned analysis are described. Distribution of voting results is part of the planned intervention. \\ \addlinespace[7pt]
Human Enhancement Technologies: A Survey Experiment on Private Demand and Governance Preferences\newline AEARCTR-0016972\newline 30 / 30 / 30 & Registry first-registered field: 8 Oct 2025. Recorded registry version 1.0 published: 13 Oct 2025. & Describes a short planned Prolific pilot. \\ \addlinespace[7pt]
Training Novices to Think\newline AEARCTR-0017124\newline 10 / 10 / 10 & Registry first-registered field: 28 Oct 2025. Recorded registry version 2.0 published: 22 Mar 2026. & Intervention, outcomes and planned analysis are described. \\ \addlinespace[7pt]
Contact and the Price of Prejudice\newline AEARCTR-0018089\newline 7 / 7 / 7 & Registry first-registered field: 13 Mar 2026. Recorded registry version 1.0 published: 16 Mar 2026. Plan dated 13 Mar 2026. & Numerical effects from cited earlier discrimination research, plus planned contact experiment. \\ \addlinespace[7pt]
Preferences for Antitrust Enforcement\newline AEARCTR-0018249\newline 46 / 46 / 46 & Registry first-registered field: 4 May 2026. Recorded registry version 1.0 published: 11 May 2026. & Survey treatment includes a description of an earlier search-engine experiment and its findings. \\ \addlinespace[7pt]
Taxpayer Behavior in the Age of AI: A Field Experiment on Property Tax Appeals\newline AEARCTR-0018430\newline 9 / 9 / 9 & Registry first-registered field: 1 May 2026. Recorded registry version 1.0 published: 6 May 2026. & Intervention delivery description and appeal instructions. \\ \addlinespace[7pt]
Selective Admissions in Nursing Homes: Evidence from a Randomized Audit Study\newline AEARCTR-0018639\newline 275 / 275 / 275 & Registry first-registered field: 13 May 2026. Recorded registry version 1.0 published: 18 May 2026. Plan dated 20 Sep 2024. & No main study effect finding identified. Explicit July 2024 pilot findings, including race differences of 15, 11 and 20 percentage points and payer/vignette patterns. \\ \addlinespace[7pt]
The Effects of SNAP Sugary Drink Restrictions on Consumption and Welfare\newline OSF-5nf3b\newline 14 / 14 / 14 & Plan dated 20 Apr 2026; registration deposited 20 Apr 2026 (98mzg). Project made public 5 Jul 2026; recorded registration embargo end 31 Jul 2026. & Historical-data design work and table/figure examples labeled as hypothetical. \\ \addlinespace[7pt]
The Effects of School Phone Bans: National Evidence from Lockable Pouches\newline OSF-8twh9\newline 35 / 35 / 0 & OSF registration deposited 24 Dec 2025. Supplied excerpt describes data work through early winter 2026. & No main study effect finding identified. Outcome data had been acquired and cleaned; researchers state outcome/design commitments preceded examination of impact estimates. \\ \addlinespace[7pt]
Do WhatsApp news channels help users navigate the information environment? A field experiment on information-related attitudes\newline OSF-fv945\newline 4 / 4 / 4 & OSF registration deposited 23 Sep 2024. Recorded embargo end: 31 Jan 2025. & Prospective design and questionnaire, including electoral facts used in the questionnaire. \\ \addlinespace[7pt]
\end{longtable}
\noindent Notes: Question counts refer to the screened Jan+, Feb+, and May+ samples. Descriptions summarize the saved text supplied to forecasters and the documented preparation steps. Dates distinguish registry registration, the recorded registry version, and the plan itself. Registry-version dates refer to the registry record. OSF deposit and public-release dates are distinguished where available.
\endgroup

\end{document}